\documentclass[12pt]{article}
\usepackage{graphicx} 
\usepackage{amsfonts,amssymb,amsmath}
\usepackage[letterpaper,top=2.92cm,bottom=3.15cm,left=1.9cm,right=1.9cm,
marginparwidth=1.75cm,headheight=0.96cm,footskip=1.5cm]{geometry}
\usepackage{hyperref}
\usepackage{xcolor}
\usepackage[titletoc,title]{appendix}
\usepackage{mathtools}
\usepackage{authblk}
\newtheorem{theorem}{Theorem}
\usepackage{commath}
\usepackage{siunitx}

\usepackage[
backend=biber,
style=ieee,
]{biblatex}
\title{A causality inspired acceleration method for the fast temporal superposition of the finite line source solutions}

\author[1]{Marc Basquens}
\author[1]{Alberto Lazzarotto}
\affil[1]{Department of Energy Technology, Division of Applied Thermodynamics and Refrigeration, KTH Royal Institute of Technology, Stockholm, Sweden}
\date{}

\begin{document}

\maketitle

\begin{abstract}
We present a novel, fast method to compute thermal interactions in solids, useful for time-dependent problems involving several sources and several time and space scales such as the ones encountered in the physics of fields of closed loop borehole heat exchangers. The new method is based on the non-history temporal superposition acceleration algorithm, but presents better performance compared to the originally proposed scheme. The main idea behind it is to leverage the propagation properties of the heat wave.
Despite the basic physical solutions of heat transfer being non-causal, it is possible to establish an influence region by fixing an acceptable error tolerance.
This allows to reduce the necessary integration regions in such a way that numerical integration is favored.
The better behaviour of the integrand arising from this approach allows us to replace the use of Bakhalov-Vasil'eva method in favor of the asymptotic method for the computation of highly oscillatory integrals that has better properties from a computational perspective in the present application. Extensive testing is presented to evaluate the robustness of the new method and to compare its performance against the originally proposed non-history method and the convolution using the FFT algorithm for a range of error tolerances. The results show that the computational cost is highly reduced for the precomputation, which includes all the computations done before starting the time-stepping scheme. The reduction is of several orders of magnitude, depending on the specific case. This cost was the bottleneck of the original non-history implementation, and reducing it in this way makes the method suitable for simulations involving hundreds of sources and hundreds of thousands of time steps that can arise in simulations of borehole fields.
\end{abstract}

\section{Introduction}

Closed-loop arrays of vertical borehole heat exchangers, coupled with heat pumps, are becoming an increasingly popular solution for meeting the heating and cooling demands of buildings while simultaneously reducing greenhouse gas emissions \cite{LUND2021101915}. To support their effective deployment, detailed numerical simulations are essential. These simulations enable engineers to optimize system design, assess long-term sustainability, and minimize operational risks.

A major challenge in modeling such geothermal systems lies in the multiple spatial and temporal scales required to describe the thermal problem, especially for large borefields. Boreholes typically have diameters ranging from \SI{100}{\milli\meter} to \SI{150}{\milli\meter}, yet can extend hundreds of meters in depth \cite{ASHRAE}. Their slender geometry, combined with the close spacing -- often just a few meters apart -- creates a complex thermal problem characterized by multiple interacting scales: from centimeters to hundreds of meters, and from minutes to decades \cite{LI2015178, ClaessonJaved2011}.
To address this complexity, simulation approaches often rely on the superposition in time and space of thermal effects modeled as finite line sources associated with each borehole \cite{CIMMINO2013401, ClaessonJaved2011}. This family of methods has become a widely used technique for borefield modeling in engineering applications, as it balances computational efficiency with the need for physical accuracy across the many scales involved \cite{LI2015178}. However, for long-term, high-resolution simulations -- such as hourly simulation over decades -- the computational cost of time superposition can quickly become infeasible unless acceleration methods are employed. This is due to the fact that a naïve implementation requires summing the influence of every past time step at each current step, leading to a computational complexity that grows quadratically with the number of time steps $(\mathcal{O}(N_t^2))$. To overcome this issue, several strategies have been developed to reduce the cost of time superposition \cite{MARCOTTE2008651, MitSpi2019, Bernier2004, ClaJav2012}. Among them, the non-history dependent method \cite{lamarche2007, lamarche2009} is particularly notable for its computational efficiency and its time-marching structure, which makes it well-suited for integration into broader system-level simulations.



In the recent paper \cite{non-history}, the non-history method was specialized to heat transfer in the particular cases of point source, line source to point target, and line source to line target which are the relevant geometries for borefields applications. Using this method allows for a linear computational complexity in the number of simulation time steps, making it particularly suitable for either finer temporal discretizations or longer simulations. 

While the temporal acceleration aspect of the method works as expected, the algorithm proposed in \cite{non-history} has a caveat -- the precomputation. The precomputation has the purpose of optimizing the performance by computing time-independent terms ahead of time.
In the line to point case, it involves integrals of complex exponentials and Bessel functions -- hence, oscillatory functions. 
The strategy in \cite{non-history} to handle such integrals is to use the Bakhalov-Vasil'eva integration method \cite{bakhvalov} in conjunction with standard Gauss quadrature techniques \cite{kronrod}.
However, the concrete approach presents two fundamental issues: lack of a priori error control and costly precomputation computation.
If one is interested in simulating borefields, where the interactions (and hence, the number of precomputations) scales as the square of the number of boreholes, the proposed method quickly becomes unfeasible.

In this paper, we address both problems of absence of error control and costly precomputations for the non-history method put forward in \cite{non-history}.
The key realization is to consider a notion of causality in the problem: one can omit from the computation the loads in the source that have not had time yet to influence the target up to a certain amplitude. This approach limits the ranges of the integrals, making them tractable by means of more performant integration methods.

The paper is structured as follows. In Section \ref{section:methodology} we present the core idea of using the propagation of heat waves and how it fits in the non-history approach and we propose the `blocks' method. In Section \ref{section:evaluation} we present the technical details of how to implement the method. In Section \ref{section:example} we illustrate through an example the numerical benefits of the method with regards to the involved numerical intergrals.  
In Section \ref{section:results} we analyze the accuracy and the performance of the proposed method.
We end the paper with the conclusions in Section \ref{section:conclusions}.
The technical derivations that, while critical to the method, would otherwise disrupt the reading flow of the paper, have been relegated to the Appendices.

\section{The blocks method} 
\label{section:methodology}

\subsection{Background}
We briefly recall the main points of the non-history method \cite{non-history, lamarche2007,lamarche2009}.
The temperature field generated by a source can be expressed as
\begin{align} 
\label{eq:temperature_from_F}
     T(\tilde{t}) - T_0 = \frac{1}{k_g} \frac{r_b^2}{\alpha} \int_0^\infty F(\zeta, \tilde{t}) \ \mathrm{d}\zeta \ ,
\end{align}
where the function $F$ is defined as
\begin{align}
\label{eq: F_definition}
    F(\zeta, \tilde{t}) = \int_0^{\tilde{t}} \, e^{-\zeta^2 (\tilde{t}-\tilde{\tau})} q^\prime(\tilde{\tau}) v(\zeta) \ \mathrm{d} \tilde{\tau}  \ ,
\end{align}
$v$ is the characteristic function of the system, closely related to its geometry, $q^\prime$ is the load (per unit length, when it applies), and  $\tilde{t} = \frac{\alpha}{r_b^2} t$ is the non-dimensional time.
Discretizing the load $q^\prime$ in time steps (so that it becomes a piecewise constant function), one can solve the integral in Equation \eqref{eq: F_definition} to obtain
\begin{align}
\label{eq:F_no_recursive}
    F \left(\zeta, \left(n+1\right)\Delta \tilde{t}\right) = \left( 1 - e^{-\zeta^2 \Delta \tilde{t}} \right) \frac{v(\zeta)}{\zeta^2} \sum_{i=0}^n q^\prime \left(i \Delta \tilde{t} \right) e^{-\zeta^2 (n-i) \Delta \tilde{t}} \ .
\end{align}
The key realization of the non-history method is that the expression of $F$ can be recast into the recursive form
\begin{align}
\label{eq:F_recursive}
    F(\zeta, (n+1)\Delta \tilde{t}) = e^{-\zeta^2 \Delta \tilde{t}} F(\zeta, n \Delta\tilde{t}) + \tilde{q}^\prime(\tilde{t}) \left(1-e^{-\zeta^2 \Delta \tilde{t}}\right) \frac{v(\zeta)}{\zeta^2} \ ,
\end{align}
with $F(\zeta, 0) = 0$. This fact allows us to set up a marching scheme to compute the response of a load variable in time, which at each new time step only requires the most recent value of the load (rather than the whole history of loads necessary to compute the convolution). 

If the geometry factor $v$ is oscillatory, as happens in the usual cases of interest with both point and line sources and targets, then the integral \eqref{eq:temperature_from_F} is also oscillatory. The method presented in \cite{non-history} to tackle this problem combines  Gaussian quadratures and the Bakhalov-Vasil'eva integration method (outlined in Appendix \ref{appendix:bakhalov}) to treat such integrals.
Then, \eqref{eq:temperature_from_F} can be easily computed with
\begin{align*}
    T(n \Delta\tilde{t}) - T_0 = \tilde{q}^\prime(n \Delta\tilde{t}) K + \sum_{s=0}^n \lambda_s  F(\zeta_s, \tilde{t}) \ ,
\end{align*}
where $K$ and $\lambda$ are constant throughout the simulation. For this reason, we refer to them as the precomputation.

The main advantage of this approach is its linear computational complexity in the number of time steps, making it suitable for simulations that are long or that have fine time steps.
However, it had two issues. First, due to the oscillatory nature of the integrals, it was costly to compute $\lambda$, even if it had to be done only once. But even worse, in the line source cases, the large variation in distance between source and target along the line integration also introduces numerical difficulties. 
The second problem was the lack of error control. In the paper, the number of integration points is used as a parameter to adjust the accuracy, but no \textit{a priori} way to estimate the error was provided. Instead, the error incurred was measured as a function of the number of discretization points. This situation is the opposite of the ideal case, where we set an error tolerance and then choose the number of integration points.

In this paper, we present an alternative way of numerically evaluating \eqref{eq:temperature_from_F} that solves the two problems of the method in \cite{non-history}.
The core idea of this new method is the assumption that loads that are applied now by a source will only be noticed at another target location after a certain amount of time. This requirement stands on the assumption that heat transfer is causal, i.e., it takes some time for the heat wave front to travel from one place to another. While this may seem intuitively obvious, recall the fundamental solution of the heat equation, describing the temperature profile generated by a point heat source, instantaneously released at $t=0$:
\begin{align*}
    T(r, t) - T_0 = \frac{\alpha}{k_g} \frac{1}{\left( 4 \pi \alpha t\right)^{3/2}} e^{-\frac{r^2}{4\alpha t}} \ .
\end{align*}
This expression tells us that the temperature $T$ at any time $t>0$ after the heat is released is non-zero \textit{at any point in space}. This makes the solution physically non-causal \cite{Christov_Jordan, Giusti, Herrera}, which conflicts with our knowledge of physics. However, examining the behaviour of the fundamental solution, we realize that its value is small \textit{until the heat wave arrives} at which point the temperature change becomes quickly noticeable.
Therefore, the tension between this non-causal model and our causal experience of reality can be alleviated by choosing a tolerance $\epsilon$ and adopting the posture that any value below $\epsilon$ can be neglected.
This mimics the effect of causality, and since such small values below $\epsilon$ can not be reliably measured in practice, this results in a good working model.

\subsection{A causal interpretation of the non-history method}
The strategy we propose is to take advantage of the notion of causality. More precisely, by using the geometric information of the system, we can determine how much time $\tau_{i\rightarrow j}$ it takes for a particular source $i$ to influence a particular target $j$ (up to an error $\epsilon$). Then, at time $\tau$, loads occurring at the source $i$ at times $\tau > t > \tau - \tau_{i\rightarrow j}$ can be disregarded when computing the interaction $i \rightarrow j$ (up to error $\epsilon$).
In the non-history scheme \eqref{eq:F_recursive}, this can be translated into the fact that, if we ignore the last $N$ loads (this is, setting it to $0$ for those last $N$ time steps), then as can be seen from either Equation \eqref{eq:F_no_recursive} or \eqref{eq:F_recursive}, the value of $F$ is
\begin{align*}
     F(\zeta, n \Delta \tilde{t}) =  e^{-\zeta^2 N \Delta \tilde{t}}  F(\zeta, (n - N) \Delta\tilde{t}) \ .
\end{align*}
By doing this, $F$ picks up a suppressing Gaussian factor with respect to the last value of $F(\zeta, (n-N) \Delta\tilde{t})$ where the load had causal influence. 
As we will show, leveraging this fact simplifies the numerical integration in $\zeta$ in Equation \eqref{eq:temperature_from_F}.

\begin{figure}[h!]
    \centering
    \includegraphics[width=0.85\linewidth]{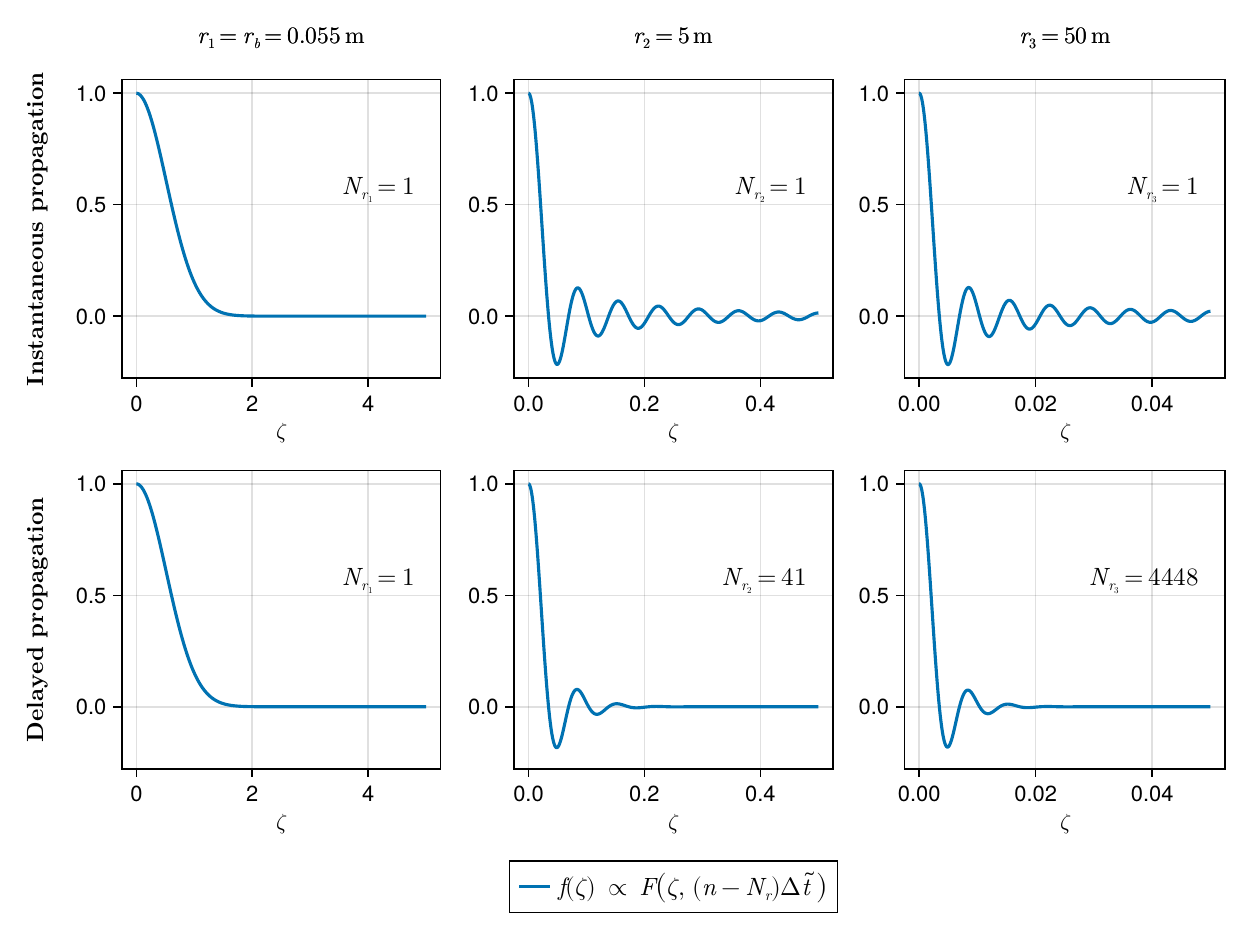}
    \caption{Illustration of the effect of considering causality on the integrand function arising in the non-history scheme for the point source solution. 
    The scale of the coordinate $\zeta$ for the three values of the radius has been chosen to optimize the illustration of the effect of causality in the three cases. For the calculation it was considered for a $\Delta t = \,$ \SI{1}{\hour}, a thermal diffusivity $\alpha = $\SI{1.44e-6}{\square\meter\per\second}, a thermal conductivity $k = $\SI{3}{\watt\per\meter\per\kelvin} and a tolerance $\epsilon = 10^{-16}$.
    }
    \label{fig:decay_example}
\end{figure}

A first example of the working principle of the method in action is illustrated in {Figure \ref{fig:decay_example}. The figure shows the integrand functions arising from the method in the case of an impulse response due to a point source evaluated at three distances \SIlist{0.055; 5; 50}{\meter}, for a $\Delta t = \,$ \SI{1}{\hour}, a thermal diffusivity $\alpha = \;$\SI{1.44e-6}{\square\meter\per\second} and a conductivity $k = \;$\SI{3}{\watt\per\meter\per\kelvin}. 
The first row of plots shows the integrand function resulting from an impulse applied one step in advance ($N_r=1$). This subset of plots is referred to as \emph{instantaneous propagation} and do not take into account causality. In these cases, the procedure does not recognize that after one step the response at \SIlist{5;50}{\meter} is negligible. Despite the integral being essentially zero, the integrand is a slowly decaying oscillating function that requires special treatment for numerical integration  \cite{non-history}.
Conversely, the second row of plots, referred to as \emph{delayed propagation}, considers the evaluation of the integral resulting from an impulse applied $N_r$ steps ahead of the evaluation time, where $N_r \, \Delta\tilde{t}$  corresponds to the time when the response at distance $r$ is non-negligible up to a tollerance $\epsilon$.
The resulting integrands are decaying much faster compared to the previous cases, and only a limited number of oscillation periods takes place before the function approaches zero, showing that the increase of the oscillation rate due to the increase of the distance $r$ is compensated by a faster decaying gaussian induced by a longer delay $N_r \Delta t$. These observations suggest that for the point source case, if causality is considered, the integral is well behaved and Gauss integration is adequate for integration along $\zeta$. This first qualitative result is a step in the right directions to overcome the limitations posed by the original non-history method. 
As it will be shown later in the paper, the exact same mechanism can be refined and exploited to effectively compute temperature responses for the cases of line source to point target, and line source to line target.

Let us develop this idea further to consider arbitrary sequences of load pulses.
We define \textit{the source function} as the Gaussian smoothing of the load $q^\prime$ (present in the original formulation \eqref{eq:F_no_recursive}):
\begin{align*}
    s (\zeta, n \Delta \tilde{t}) \vcentcolon = \sum_{i=0}^{n} q^\prime(i \Delta \tilde{t}) \ e^{-\zeta^2 \left( n - i \right) \Delta \tilde{t}} \ .
\end{align*}
This time-dependent weighted load contains all the temporal information of the load for a given source. 
The crucial property of $s$ is that, unlike the convolution with the step response, its value at the next time step can be computed from the value at the current time step:
\begin{align}
\label{eq:source_recursive}
    s\left(\zeta, (n+1)\Delta \tilde{t} \right) = q^\prime\left((n+1)\Delta\tilde{t}\right) + e^{-\zeta^2 \Delta\tilde{t}} s\left(\zeta, n \Delta\tilde{t}\right) \ ,
\end{align}
rather than needing the whole sequence of loads for each new evaluation.

In order to make use of the causality observation, we split $s$ into several parts that we will call blocks. Each block contains a certain range of time steps, determined by the integers $\{ N_k \}_{0\leq k \leq K}$, 
and are defined such that, at the $n$-th timestep, the $k$-th block spans the period between the time steps $n-N_{k}$ and $n - N_{k-1}$.
We take as a convention that $N_0 = 0$ and $N_K = \infty$, and that loads of negative times are $0$.
Then, we define the source function of the $k$-th block as
\begin{align}
\label{eq:source_block}
    s_k \left(\zeta, n \Delta \tilde{t} \right) \vcentcolon = \sum_{i=n-N_k+1}^{n-N_{k-1}} q^\prime(i \Delta \tilde{t}) e^{-\zeta^2 \left( n - i \right) \Delta \tilde{t}} \ .
\end{align}
A visual representation of the blocks $s_k$ is shown in Figure \ref{fig:blocks}.

The blocks segment the time domain, but as a consequence of the causality discussion, each block is also effectively associated with a certain region in space surrounding the source. For a given block, it is only inside the corresponding region where interactions are relevant, and interactions outside of it can be disregarded. 
The full solution is then obtained when we sum over all the blocks, as expected:
\begin{align*}
     s (\zeta, n \Delta \tilde{t}) = \sum_{k=1}^K s_k \left(\zeta, n \Delta \tilde{t} \right) \ .
\end{align*}

\begin{figure}[h!]
    \centering
    \includegraphics[width=\linewidth]{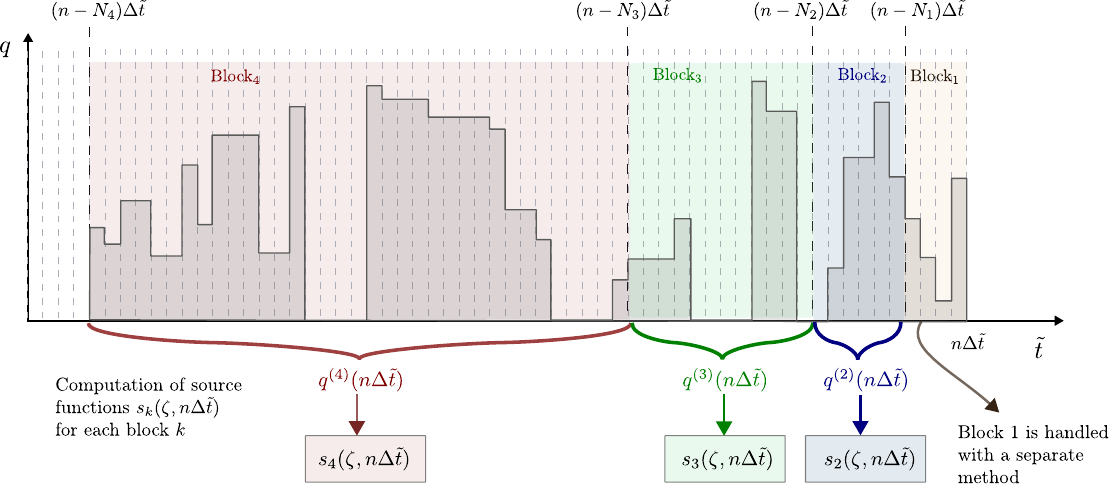}
    \caption{Visual representation of the blocks $s_k$. The blocks partition the load history, with the first block $s_1$ consisting of the most recent loads, and the last block $s_K$ consisting of the most distant in the past loads. The block $s_k$ is delimited by $N_k, N_{k-1}$, the number of time steps in the past from the current one. Since the blocks are fixed, as time passes, the loads flow from each block to the next, eventually accumulating at the last block $s_K$.}
    \label{fig:blocks}
\end{figure}

In order to use the blocks in a marching scheme, as intended in the non-history approach, we must derive an expression for the next time step of $s_k$ based on the previous one, similarly to \eqref{eq:source_recursive}:
\begin{align}
\label{eq:recursive_sk}
\begin{split}
    s_k \left(\zeta, n \Delta \tilde{t} \right) = e^{-\zeta^2 \Delta \tilde{t}} s_k \left(\zeta, (n-1) \Delta \tilde{t} \right) &+ q\left((n-N_{k-1}) \Delta \tilde{t} \right) e^{-\zeta^2 N_{k-1} \Delta \tilde{t}}  \\ 
    &- q\left((n-N_k) \Delta \tilde{t}\right) e^{-\zeta^2 N_{k} \Delta \tilde{t}} \ .
\end{split}
\end{align}
This recurrence can be interpreted as follows. At every time step $n$, each block $k$ ``receives" a load (represented by the second term in \eqref{eq:recursive_sk}), which was produced by the source at the time step $n-N_{k-1}$. At the same time, each block ``releases" a load (represented by the third term in \eqref{eq:recursive_sk}), which was produced by the source at the time step $n-N_k$. With the exception of the first and the last blocks, the load that the $k$-th block ``releases" is the one received by the $(k+1)$-th block.
The first block $k=1$ receives the load currently produced by the source, and the last block $k=K$ accumulates all the trailing loads, increasing the amount of loads to take into account at each time step.
Lastly, the effect of all the loads previously emitted by the source fades as time passes (represented by the first term in \eqref{eq:recursive_sk}).

Since we are eventually interested in simulating the evolution of several sources which potentially also act as targets, there are some remarks that are useful to keep in mind.
First, as we have already mentioned in the discussion of causality in the heat transfer process, the number of most recent timesteps that can be safely disregarded in the interaction between a source $i$ and a target $j$ depends on both the geometry of $i$ and $j$ and the error tolerance $\epsilon$ that we are willing to incur. We denote this amount by $N_{i \rightarrow j}(\epsilon)$.
Note that $N_{i \rightarrow i}(\epsilon) = 0$.
The computation of $N_{i \rightarrow j}(\epsilon)$ is discussed in Section \ref{section:blocks}.

Second, each interaction between each pair of source $i$ and target $j$ is associated with the corresponding geometry factor $v_{i\rightarrow j}$. However, each source $i$ has associated its own source function $s^i$, that only depends on the loads occurring at $i$.
Rather than updating at each time step the function $F$ for each interaction, whose amount scales with the square $n_s^2$ of the number of sources $n_s$, we can simply update the $n_s$ functions $s^i$.
Then, to recover the temperature difference generated by a source $i$ at a given target $j$, we just need to integrate the corresponding geometry factor $v_{i\rightarrow j}$ multiplied by $s^i$.

For a particular interaction between a source $i$ and target $j$, we only need to keep the blocks of the source that are relevant according to $N_{i \rightarrow j}(\epsilon)$. We define
\begin{align*}
    K_{i \rightarrow j}(\epsilon) = \max_{k} \{ k \ \rvert \ N_k \leq N_{i \rightarrow j}(\epsilon) \} \ ,    
\end{align*}
as the first block that needs to be taken into account for this particular interaction.
Then, the temperature difference generated by a source $i$ at a target $j$ is given by
\begin{align*}
    T_{i\rightarrow j}(n \Delta \tilde{t}) - T_0 = \frac{1}{k_g} \frac{r_b^2}{\alpha}\int_0^\infty  \left( 1 - e^{-\zeta^2 \Delta \tilde{t}} \right) \frac{v_{i\rightarrow j}(\zeta)}{\zeta^2} \sum_{k= K_{i \rightarrow j}(\epsilon)}^K s^i_k \left(\zeta, n \Delta \tilde{t} \right) \ \mathrm{d} \zeta \ .
\end{align*}

In practice, we want to perform the integrals first and then the sum in $k$:
\begin{align}
\label{eq:block_integral}
    T_{i\rightarrow j}(n \Delta \tilde{t}) - T_0 = \frac{1}{k_g} \frac{r_b^2}{\alpha}\sum_{k= K_{i \rightarrow j}(\epsilon)}^K \int_0^\infty  \left( 1 - e^{-\zeta^2 \Delta \tilde{t}} \right) \frac{v_{i\rightarrow j}(\zeta)}{\zeta^2}  s^i_k \left(\zeta, n \Delta \tilde{t} \right) \ \mathrm{d} \zeta \ .
\end{align}
This is because of two main reasons.
First, since $s_k$ is proportional to $e^{-\zeta^2 N_{k-1} \Delta \tilde{t}}$ (according to \eqref{eq:source_block}), each of the integrands is suppressed (with the exception of $k=1$), reducing the amplitude of the oscillations and the effective numerical cutoff for the $\zeta$ integral. 
Second, since each block represents a finite (except for $k=K$) and limited time frame, due to causality, not all points in the source lines will affect the whole target. Hence, we can approximate $v_{i\rightarrow j}(\zeta)$ by a block-dependent function $v_{i\rightarrow j, k}(\zeta)$, in which the integration is performed over a smaller region, where the integrand does not vary as much. Details are provided for each separate case in Section \ref{section:evaluation}. 
Both effects contribute towards making the numerical integration faster.

\section{Numerical evaluation}
\label{section:evaluation}

We dedicate this section to the numerical evaluation of \eqref{eq:block_integral} by discussing three numerical aspects left to clarify in Section \ref{section:methodology}: how to choose the blocks, how to compute the geometry factors, and how to perform the $\zeta$ integral.
This Section is structured as follows.
In Section \ref{section:blocks}, we choose the block limits $N_k$ based on the geometry of the system and an error tolerance.
In Section \ref{section:v_evaluation}, we evaluate the geometry factor in an efficient way for each geometrical case.
In Section \ref{section:discretization}, we describe the procedure to obtain quadrature weights and nodes to numerically compute the integral in $\zeta$ that are usable throughout the simulation.

\subsection{Choice of blocks}
\label{section:blocks}
While from the discussion in Section \ref{section:methodology} there are no restrictions in principle to choose the block limits $N_k$, there are some choices that yield better performance than others.

The fastest oscillations of the integrand in \eqref{eq:block_integral} have a frequency proportional to the maximum distance in a block, as we show in \eqref{eq:v_approx}, later in this section.
Therefore, we want to limit the maximum distance that will be causally affected by a source within each block.

The first block is determined by $N_0 = 0$ and $N_1$ equal to the minimum time at which the thermal influence of the source affects the closest target up to error $\epsilon$. Then, the subsequent blocks are chosen by limiting the maximum distance relative to the minimum distance in each block. To translate into equations, let $f(r, t)$ be the function that evaluates the error commited by not evaluating the effects of the sources at a distance $r$ at time $t$ (note that this is generic notation, the specific spatial variables on which $f$ depends will be made precise in each case). Then, at the block $s_k$, assuming that we want to limit the distance at $r_k$, we find $N_k$ as the highest integer satisfying
\begin{align}
\label{eq:N_r}
    f(r_k, N_k \Delta t) \leq \epsilon \ .
\end{align}
We have that $r_1 = r_{ij}$, where $r_{ij}$ is the shortest distance between source and target, and we can define $r_{k+1} = p\ r_k$.
Here, $p$ is an arbitrary choice, guided by the difficulty introduced by higher frequencies in the oscillatory integral. A value that yields good results is $p=3$. 

Note that we can compute $N_{i \rightarrow j}(\epsilon)$ (the number of most recent time steps that we can disregard in the interaction $i\rightarrow j$ up to error $\epsilon$) by solving
\begin{align}
\label{eq:N_r_eps}
    f(r_{ij}, N_{i \rightarrow j}(\epsilon)\Delta t) \leq \epsilon \ .
\end{align}

We now discuss the suitable functions $f$ in Equation \eqref{eq:N_r} for point and line sources and targets. Figure \ref{fig:source-geometries} provides a graphical representation of all three investigated cases along with the description of the geometric parameters that are necessary for computation. 

\begin{figure}[h!]
    \centering
    \includegraphics[width=0.7\linewidth]{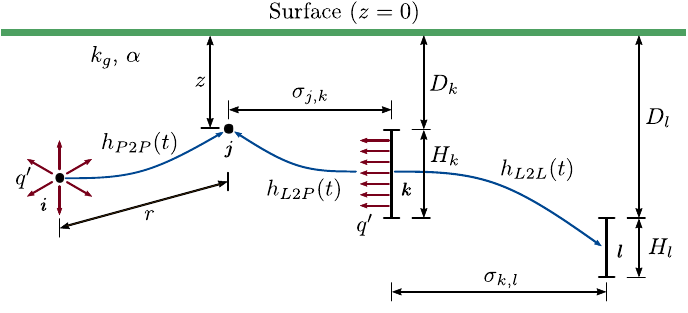}
    \caption{Geometry of sources and targets.}
    \label{fig:source-geometries}
\end{figure}
\subsubsection{Point to point}

Since the step response of a point source is
\begin{align*}
    h^\text{P2P}(r, t) = \frac{\text{erfc } \frac{r}{\sqrt{4\alpha t}}}{4 \pi k_g r} \ ,
\end{align*}
by disregarding a single point source at a distance $r$ we commit an error $h^\text{P2P}(r, t)$.
Therefore, $f = h^\text{P2P}$.

\subsubsection{Line to point}

This case is slightly more complicated than the previous one. Since the source is a line, the thermal effect of each of its points does not arrive at the same time at the target point. For small distances of the point to the line, or small times, only a portion of the line is thermally affecting the target point, up to a certain tolerance.

The complete step response of the line source is
\begin{align*}
    h^\text{L2P}(t) &=\frac{\alpha}{k_g} \int_{D_i}^{D_i+H_i} \int_{0}^{t} \frac{1}{(4\pi\alpha \tau)^{\frac{3}{2}}} e^{-\frac{\sigma^2 + (z_j-z^\prime)^2}{4\alpha \tau}}  \mathrm{d}\tau \ \mathrm{d}z^\prime  \\
    &= \frac{1}{4 \pi k_g}\int_{\frac{1}{\sqrt{4\alpha t}}}^\infty  \frac{e^{-\sigma^2 s^2}}{s} \Bigg( \text{erf } \Big( s ( z_j-D_i) \Big)  - \text{erf } \Big( s ( z_j-D_i-H_i) \Big) \Bigg)\ \mathrm{d}s \ ,
\end{align*}
however, if we restrict the integral to the causally relevant region of the line, which we parametrize by its center $z$ and length $2h$, the restricted step response becomes
\begin{align*}
    \bar{h}^\text{L2P}(h, t) &=  \frac{\alpha}{k_g} \int_{\max\{z - h, \ D\}}^{\min\{z + h, \ D+H\}} \int_{0}^{t} \frac{1}{(4\pi\alpha \tau)^{\frac{3}{2}}} e^{-\frac{\sigma^2 + (z-z^\prime)^2}{4\alpha \tau}}  \mathrm{d}\tau \ \mathrm{d}z^\prime  \\
    &= \frac{1}{2 k_g \pi^{\frac{3}{2}}} \int_{\frac{1}{\sqrt{4\alpha t}}}^{\infty}   \int_{\max\{z - h, \ D\}}^{\min\{z + h, \ D+H\}}  e^{-s^2(\sigma^2 + (z-z^\prime)^2)} \mathrm{d}z^\prime \mathrm{d}s \\
    &=  \frac{1}{4 \pi k_g}\int_{\frac{1}{\sqrt{4\alpha t}}}^\infty  \frac{e^{-\sigma^2 s^2}}{s} \Bigg( \text{erf } \Big( s \min \{ z-D, h \} \Big)  - \text{erf } \Big( s \max \{ z-D-H, -h \} \Big) \Bigg)\ \mathrm{d}s \ .
\end{align*}
Clearly, the error commited is the difference between the complete and the restricted step responses.
Then, for a given time $t$, we use equation \eqref{eq:N_r} to choose the value $h$ that yields an error less than $\epsilon$ by defining
\begin{align*}
     f(h, t) = \Big\lvert \bar{h}^\text{L2P}(h, t) - h^\text{L2P}(t) \Big\rvert\ .
\end{align*}

\subsubsection{Line to line}
This case is similar to the previous one, but with a difference. Since the target is also a line, each of its points has a different region of the source line that is affecting it up to the tolerance $\epsilon$. 

The complete step response using both source and target lines is:
\begin{align*}
    h^\text{L2L}(t) &= \frac{\alpha}{k_g H^\prime} \ \int_{D_j }^{D_j +H_j }  \int_{ D_i}^{ D_i+H_i} \int_0^t \frac{1}{(4 \pi \alpha \tau)^\frac{3}{2}} e^{-\frac{\sigma^2+ (z_j-z^\prime)^2}{4\alpha \tau}} \mathrm{d}\tau \ \mathrm{d}z^\prime \ \mathrm{d}z
    \\ &=\frac{1}{4 \pi k_g H'}\int_{\frac{1}{\sqrt{4\alpha t}}}^\infty \frac{e^{-\sigma^2 s^2}}{s^2}\Bigg(  \text{ierf } \Big( s (D_j + H_j - D_i)\Big) + \text{ierf } \Big( s ( D_i + H_i - D_j) \Big) \\
    &\hspace{4.3cm}-\text{ierf } \Big( s (D_j-D_i) \Big) - \text{ierf } \Big( s (D_j+H_j-D_i-H_i) \Big)    \Bigg)   \mathrm{d}s \ ,
\end{align*}
while the restricted response is
\begin{align*}
    \bar{h}^\text{L2L}(h, t) &= \frac{\alpha}{k_g H_j} \ \int_{D_j }^{D_j +H_j }  \int_{\max \{z - h, \ D_i \}}^{\min \{z + h, \ D_i+H_i\}} \int_0^t \frac{1}{(4 \pi \alpha \tau)^\frac{3}{2}} e^{-\frac{\sigma^2+ (z_j-z^\prime)^2}{4\alpha \tau}} \mathrm{d}\tau \ \mathrm{d}z^\prime \ \mathrm{d}z
    \\ &=\frac{1}{4 \pi k_g H_j}\int_{\frac{1}{\sqrt{4\alpha t}}}^\infty \Biggr\{ \frac{e^{-\sigma^2 s^2}}{s^2}\Bigg[  \text{ierf } \Big( s \min \{ D_j + H_j - D_i, h \}\Big) + \text{ierf } \Big( s \min \{ D_i + H_i - D_j, h \} \Big) \\
    &\hspace{5cm}-\text{ierf } \Big( s (D_j-D_i) \Big) - \text{ierf } \Big( s (D_j+H_j-D_i-H_i) \Big)    \Bigg] \\
    &\hspace{1cm}+\frac{e^{-\sigma^2 s^2}}{s} \text{erf } (sh)\Bigg[  \min \{ H_j, D_j+H_j-D_i-h \} +   \min \{ H_j, D_i+H_i-D_j-h \}  \Bigg] \Biggr\} \mathrm{d}s \ .
\end{align*}
Again, the error commited is the difference between the complete and restricted step responses, hence, for a given time $t$, we find the necessary $h$ via
\begin{align*}
     f(h, t) = \Big\lvert \bar{h}^\text{L2L}(h, t) - h^\text{L2L}(t) \Big\rvert \ ,
\end{align*}
and \eqref{eq:N_r}.

\subsection{Evaluation of $v$}
\label{section:v_evaluation}
Next, we turn to the evaluation of the geometry factors for each of the cases.
\subsubsection{Point to point}
\label{section:point_to_point}
In this case, as shown in \cite{non-history}, the geometry factor is 
\begin{align}
\label{eq:v_P2P}
    \frac{v^{P2P}_{i\rightarrow j}(\zeta)}{\zeta^2} = \frac{\alpha}{2 \pi^2 r_b^3 } \frac{\sin{\tilde{r} \zeta}}{\tilde{r} \zeta} \ ,
\end{align}
where $r$ is the distance between the source $i$ and the target $j$, $r_b$ is the borehole radius, and $\tilde{r} = r/r_b$.

\subsubsection{Line to point}
\label{section:line_to_point}
The geometry factor of the line to point case is the integral
\begin{align}
\label{eq:v_L2P}
    \frac{v^{L2P}_{i\rightarrow j}(\zeta)}{\zeta^2} = \frac{\alpha}{2 \pi^2 r_b^3 } \int_{D_i}^{D_i+H_i} \frac{\sin{\tilde{r}(z)\zeta}}{\tilde{r}(z) \zeta} \ \mathrm{d} z  \ ,
\end{align}
where $D_i, H_i$ are the depth and length of the source line $i$, $\tilde{r}(z) = \frac{1}{r_b}\sqrt{\sigma^2 + (z - z_j)^2}$ where $\sigma$ is the transverse distance between the vertical line $i$ and the target $j$, and $z_j$ is the $z$ coordinate of the target. 

The direct evaluation of \eqref{eq:v_L2P} is challenging due to the oscillatory nature of the integrand, however, the blocks structure can be exploited to simplify it. The intuitive idea of the derivations that follow is that within a given block, the range of influence of the source is bounded. In the particular case of a line source and target point, this means that, for the most recent blocks, only part of the line is effectively influencing the target point and the rest can be discarded. This reduces the effective integration region in those blocks, making the numerical integration simpler. 

\paragraph{Reduced domain of influence}
As discussed in Subsection \ref{section:blocks}, depending on the considered time scale, some points in the line source will have a very small effect on the target point due to their relative distance, making the effective source a smaller line. 
This means that, in each block $k$, we can actually approximate $v_{i\rightarrow j}(\zeta)$ in Equation \eqref{eq:block_integral} by $\bar{v}_{i\rightarrow j; k}(\zeta)$, which only takes into account a smaller region of the line source parametrized by $h$:
\begin{align}
\label{eq:v_L2P_bar}
    \frac{\bar{v}^{L2P}_{i\rightarrow j; k}(\zeta)}{\zeta^2} =  \frac{\alpha}{2 \pi^2 r_b^3 } \int_{\max\{z_j - h, \ D_i\}}^{\min\{z_j + h, \ D_i+H_i\}} \frac{\sin{\tilde{r}(z^\prime)\zeta}}{\tilde{r}(z^\prime) \zeta} \ \mathrm{d} z^\prime  \ .
\end{align}
In order to determine what value of $h$ is suitable for a certain block, we examine the error that we commit in evaluating the step response by using this reduced interval instead of the full integral as explained in Subsection \ref{section:blocks}.

The main effect of the reduction of the integration domain is that the numerical integral \eqref{eq:v_L2P_bar} becomes easier than its counterpart that has the whole line as the integration region.
Note that the effect of the reduction is only relevant for the first blocks in configurations with small transverse distances $\sigma$ between the source and the target (in other cases \eqref{eq:N_r} will just yield $h > H$, which spans the whole line). 
However, precisely in these cases, if we considered the whole line, $r(z^\prime)$ would vary quite a lot along it (recall the slender geometry of the problem).
This would introduce a mix of slower and faster oscillation modes which would make the numerical integration harder. 
By restricting the integrated region, the faster oscillation modes effectively disappear.
We show the details of this intuition in the following.

\paragraph{Integral manipulation}
Depending on the particular configuration and symmetries of the system, the function $r(z)$ may not be injective along the line. This means that the integrand of \eqref{eq:v_L2P_bar} will attain some values more than once. It will be convenient to rewrite the integral so that this does not occur, by using the procedure shown in Appendix \ref{appendix:line_to_point}:
\begin{align}
\label{eq:v_L2P_r}
    \frac{\bar{v}^{L2P}_{i\rightarrow j; k}(\zeta)}{\zeta^2} =  \frac{\alpha}{2 \pi^2 r_b^3 } \int_{\max\{z_j - h, \ D_i\}}^{\min\{z_j + h, \ D_i+H_i\}} \frac{\sin{\tilde{r}(z^\prime)\zeta}}{\tilde{r}(z^\prime) \zeta}  \ \mathrm{d} z^\prime  = \frac{\alpha}{2 \pi^2 r_b^2 }\frac{1}{\zeta}\int_{\tilde{r}_{1}}^{\tilde{r}_{3}}   \frac{\sin{\tilde{r} \zeta}}{\sqrt{\tilde{r}^2 - \tilde{\sigma}^2} } \alpha_L(\tilde{r}) \ \mathrm{d} \tilde{r} \  ,
\end{align}
where according to  \eqref{eq:l2p_r_alpha} and \eqref{eq:l2p_r_limits},
\begin{align}
\label{eq:ltp_alpha}
    \alpha_L(\tilde{r}) = 
    \begin{cases}
        2 \ \text{if  } \tilde{r}_1 \leq \tilde{r} \leq \tilde{r}_2 \\
        1 \ \text{if  } \tilde{r}_2 \leq \tilde{r} \leq \tilde{r}_3
    \end{cases} \ ,
\end{align}
and
\begin{align*}
      \tilde{r}_1 &= 
    \begin{cases}
        \tilde{\sigma} \ , \     &\text{if} \ D_i \leq z_j \leq  D_i+H_i \\
        \text{min}\left( \ell_1, \ell_2 \right) \ , \ &\text{otherwise}
    \end{cases} \ , \\
    \tilde{r}_2 &= \text{min}\left(  \ell_1, \ell_2 \right) \ , \\ 
    \tilde{r}_3 &= \text{max}\left(  \ell_1, \ell_2 \right) \ , \\
\end{align*}
where
\begin{align*}
    \ell_1 &= \sqrt{\tilde{\sigma}^2 + \max\{- h, \ z_j -D_i-H_i\}^2/ r_b^2} \ , \\
    \ell_2 &= \sqrt{\tilde{\sigma}^2 + \min\{h, \ z_j-D_i\}^2/ r_b^2} \ .
\end{align*}
An illustration of the geometrical interpretation of these parameters is provided in Figure \ref{fig:geom_for_integral_L2P}.

\begin{figure}[h!]
    \centering
    \includegraphics[width=0.85\linewidth]{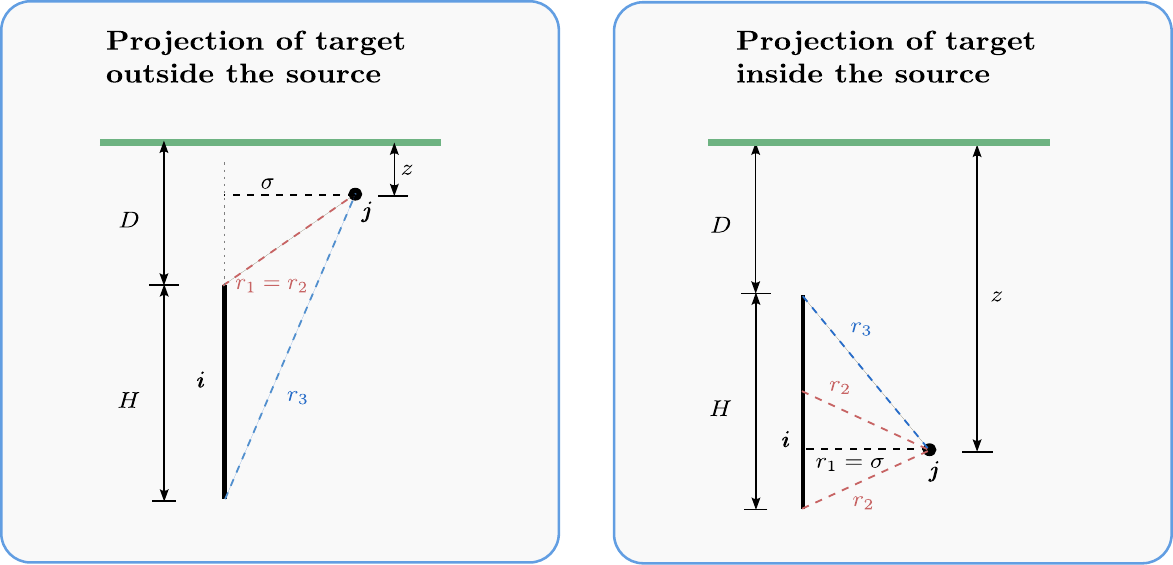}
    \caption{Illustration of two relevant cases in the estimation of the line to point function $\bar{v}^{L2P}_{i\rightarrow j; k}(\zeta)$.
    }
    \label{fig:geom_for_integral_L2P}
\end{figure}

In the form of \eqref{eq:v_L2P_r}, however, the integrand has a singularity at $\tilde{r}=\tilde{\sigma}$. Although the integral exists, if $\tilde{\sigma} \in [\tilde{r}_{1}, \tilde{r}_{3}]$, the numerical evaluation of the integral will become harder. In order to bypass this issue, we introduce yet another change of variable $\tilde{r} = \tilde{\sigma} \cosh{y}$, yielding:
\begin{align}
\label{eq:L2P_integral_cosh}
    \frac{\bar{v}^{L2P}_{i\rightarrow j; k}(\zeta)}{\zeta^2} &=\frac{\alpha}{2 \pi^2 r_b^2 }\frac{1}{\zeta} \int_{\text{arccosh }\frac{\tilde{r}_1}{\tilde{\sigma}} }^{\text{arccosh }\frac{\tilde{r}_3}{\tilde{\sigma}} } \alpha_L \left( \tilde{\sigma} \cosh{y} \right) \sin{\left(\zeta \ \tilde{\sigma} \cosh{y}\right)} \ \mathrm{d}y\nonumber \\ 
    &=\frac{\alpha}{2 \pi^2 r_b^2 }\frac{2}{\zeta} \int_{\text{arccosh }\frac{\tilde{r}_1}{\tilde{\sigma}} }^{\text{arccosh }\frac{\tilde{r}_2}{\tilde{\sigma}} }  \sin{\left(\zeta \ \tilde{\sigma} \cosh{y}\right)}\ \mathrm{d}y + \frac{\alpha}{2 \pi^2 r_b^2 }\frac{1}{\zeta} \int_{\text{arccosh }\frac{\tilde{r}_2}{\tilde{\sigma}} }^{\text{arccosh }\frac{\tilde{r}_3}{\tilde{\sigma}} }  \sin{\left(\zeta \ \tilde{\sigma} \cosh{y}\right)}
 \ \mathrm{d}y \ ,
\end{align}
where the integrands now have no singularities.

\paragraph{Numerical integration via asymptotic expansion}
Let us now discuss how to compute the oscillatory integral
\begin{align}
\label{eq:l2p_oscillatory_integral}
    I_{L2P}(a, b, \omega) = \int_a^b \sin{\left(\omega \cosh{x}\right)} \mathrm{d}x \ ,
\end{align}
that appears twice in \eqref{eq:L2P_integral_cosh}, with $\omega = \zeta \ \tilde{\sigma}$.

A simple but effective method for oscillatory integrals is the asymptotic expansion in powers of $\frac{1}{\omega}$ \cite{Iserles_Norsett, olver}, which is outlined in Appendix \ref{appendix:asymptotic}. 
In the light of \eqref{eq:L2P_integral_cosh}, we are interested in
\begin{align*}
    I^A(a, b, \omega) = \text{Im} \left( \int_a^b e^{i \omega \cosh{x}}  \mathrm{d}x\right) \ ,
\end{align*}
which corresponds to $f(x) = 1$ and $g(x) = \cosh{x}$ in \eqref{eq:asymptotic_integral}.
In this case, the functions \eqref{eq:sigma_functions} are
\begin{align}
\label{eq:sigma_functions_cosh}
\begin{split}
    \sigma_{2k}(x) &= -\frac{\cosh{x}}{(\sinh{x})^2}\sum_{p=k-1}^{2(k-1)} \frac{a^{2k}_p}{(\sinh{x})^{2p+1}} \ , \\
    \sigma_{2k+1}(x) &= \sum_{p=k}^{2k} \frac{a^{2k+1}_p}{(\sinh{x})^{2p+1}} \ ,
    \end{split}
\end{align}
where the coefficients $a^n_p$ satisfy the recurrence relation
\begin{align}
\label{eq:sigma_functions_cosh_a}
\begin{split}
    a^{2k}_p &= -  (2p+1) \ a^{2k-1}_p \ , \hspace{4.85cm}  \text{for } p=k-1, \dots, 2k-2\\
     a^{2k+1}_{p} &= (2p-1)(2p-3) \ a^{2k-1}_{p-2} +2p(2p-1) \ a^{2k-1}_{p-1} , \quad  \text{for } p=k+1, \dots, 2k-1 \\
    a^{2k+1}_k &= 2k (2k-1)\ a^{2k-1} \ , \\
    a^{2k+1}_{2k} &= (4k-1)(4k-3) \ a^{2k-1}_{2k-2} \ , \\
    \end{split}
\end{align}
and $a_0^1 = 1$.
Then, according to \eqref{eq:asymptotic_formula},
\begin{align}
\label{eq:asymptotic_formula_l2p}
    I_n^A(a, b, \omega) = \text{Im}\left( \sum_{k=1}^n \frac{1}{\left( - i \omega\right)^k} \left( 
    \sigma_k(b) e^{i\omega \cosh{b}} - \sigma_k(a) e^{i\omega \cosh{a}} \right) \right)\ .
\end{align}

\paragraph{Combining Gauss integration and asymptotic expansion}
A good feature of the asymptotic expansion is that it converges quickly for highly oscillating integrals, but its downside is that it converges poorly for small values of $\omega$. 
Moreover, in our particular case, the lower limit of integration of \eqref{eq:L2P_integral_cosh}, satisfies that $a = \text{arccosh }\frac{r_1}{\sigma} \rightarrow 0$ as $r_1 \rightarrow \sigma$. Since \eqref{eq:sigma_functions_cosh} are proportional to $\frac{1}{\sinh{x}}$, and \eqref{eq:asymptotic_formula_l2p} depends on $\sigma_k(a)$, when $r_1$ is close to $\sigma$, \eqref{eq:asymptotic_formula_l2p} will be numerically unstable.

Therefore, we cannot reliably use \eqref{eq:asymptotic_formula_l2p} for all values of $a$ and $\omega$. However note that, precisely for those cases where the asymptotic method is unstable, Gaussian quadratures are suitable, namely, for small $\omega$ (slow oscillations) and $a$ close to $0$ (the expansion has a singularity at $x=0$).
The formula for this method is
\begin{align}
\label{eq:gauss_formula_l2p}
    I^G(a, b, \omega) = \sum_{k=1}^{n_G} w_k \sin{\left(\omega \cosh{x_k}\right)} \ ,
\end{align}
where $x_k$ and $w_k$ are the nodes and weights, respectively, obtained from an adaptive quadrature algorithm, for instance the Gauss-Kronrod one.

In this way, we combine both methods \eqref{eq:asymptotic_formula_l2p} and \eqref{eq:gauss_formula_l2p} to compute the integral \eqref{eq:l2p_oscillatory_integral} in the whole necessary parameter space.
To decide on what integration method we use for each value of $a$ and $\omega$, we use the error tolerance. 
The error we commit in truncating the asymptotic series up to $n$ terms is given by \eqref{eq:asymptotic_error}, which in this case reads
\begin{align*}
    E_n^A(a, b, \omega) = \frac{1}{(-i\omega)^n} \int_a^b \sinh{x} \ \sigma_{n+1}(x) e^{i \omega \cosh{x}} \mathrm{d}x \ .
\end{align*}
Since this integral is also oscillatory and trying to evaluate it would present the same kind of challenges than the original one, we instead resort to bounding it
\begin{align*}
    \lvert E_n^A(a, b, \omega)  \rvert \leq  \Bigg\lvert \frac{1}{(-i\omega)^n} \Bigg\rvert \int_a^b \Big\lvert \sinh{x} \ \sigma_{n+1}(x) e^{i \omega \cosh{x}} \Big\rvert \mathrm{d}x = \frac{1}{\omega^n} \int_a^b \sinh{x} \ \lvert\sigma_{n+1}(x)\rvert \mathrm{d}x \ .
\end{align*}
The latter integral can be computed analytically for all $n$, however, its expression is particularly simple for $n$ odd:
\begin{align*}
    \lvert E_n^A(a, b, \omega)  \rvert \leq \frac{1}{\omega^n} \sum_{p=\frac{n-1}{2}}^{n-1}  \frac{1}{2p+1}\frac{a^{n+1}_p}{(\sinh{x})^{2p+1}} \Biggr \lvert^b_a \ =: \tilde{E}_n^A(a, b, \omega).
\end{align*}

Since we know that the asymptotic method works best when $a \gg 0$, we use it in a region sufficiently far from $0$. We find the lower limit of the interval $r^\star$ by numerically solving the equation
\begin{align}
\label{eq:l2p_rstar}
    \tilde{E}_n^A(r^\star, b, \omega) = \frac{\epsilon}{2} \ .
\end{align}

Then, we approximate integral \eqref{eq:l2p_oscillatory_integral}, up to error $\epsilon$, by means of \eqref{eq:gauss_formula_l2p}, \eqref{eq:asymptotic_formula_l2p}, and \eqref{eq:l2p_rstar} as
\begin{align}
\label{eq:l2p_final}
    I_{L2P}(a, b, \omega) \simeq  I^G(a, r^\star, \omega) + I_n^A(r^\star, b, \omega) \ ,
\end{align}
where we set the tolerance for the adaptive Gaussian quadrature to $\frac{\epsilon}{2}$.

\subsubsection{Line to line}
\label{section:line_to_line}
The geometry factor is the integral
\begin{align}
\label{eq:v_L2L}
    \frac{v^{L2L}_{i\rightarrow j}(\zeta)}{\zeta^2} = \frac{1}{H_j}\frac{\alpha}{2 \pi^2 r_b^3 } \int_{D_j }^{D_j +H_j }  \int_{D_i}^{D_i+H_i}  \ \frac{\sin{\tilde{r}(z, z^\prime)\zeta}}{\tilde{r}(z, z^\prime) \zeta} \ \mathrm{d} z \ \mathrm{d} z^\prime \ ,
\end{align}
where $D_i, H_i, D_j, H_j$ are the depths and lengths of the source line $i$ and target line $j$, respectively. Moreover, $\tilde{r}(z, z^\prime) =\frac{1}{r_b} \sqrt{\sigma^2 + (z - z^\prime)^2}$ where $\sigma$ is the transverse distance between the lines $i$ and $j$, and $\tilde{r} = r / r_b$.

Similarly to the line to point case, in each block, we can perform the integration only over the relevant region (see Section \ref{section:blocks}):
\begin{align}
\label{eq:L2L_sin}
    \frac{v^{L2L}_{i\rightarrow j}(\zeta)}{\zeta^2} = \frac{1}{H_j}\frac{\alpha}{2 \pi^2 r_b^3 } \int_{D_j }^{D_j +H_j }  \int_{\max \{z - h, \ D_i \}}^{\min \{z + h, \ D_i+H_i\}}  \ \frac{\sin{\tilde{r}(z, z^\prime)\zeta}}{\tilde{r}(z, z^\prime) \zeta} \ \mathrm{d} z^\prime \ \mathrm{d} z \  .
\end{align}

According to the results in Appendix \ref{appendix:line_to_line}, in particular Equation \eqref{eq:line_to_line_expression_strip}, the integral \eqref{eq:v_L2L} can be written in the form
\begin{align}
\label{eq:L2L_r}
\begin{split}
    \int_{D_j }^{D_j +H_j } & \int_{\max \{z - h, \ D_i \}}^{\min \{z + h, \ D_i+H_i\}}  \frac{\sin{\tilde{r}(z, z^\prime)\zeta}}{\tilde{r}(z, z^\prime) \zeta} \ \mathrm{d} z \ \mathrm{d} z^\prime \\
    &= \frac{r_b}{\zeta} \int_{\tilde{r}_1}^{\min \{\tilde{r}_4, \ \tilde{r}_h\}}  \left(\frac{\sin{\tilde{r} \zeta}}{ \sqrt{\tilde{r}^2 - \tilde{\sigma}^2} } \tilde{\alpha}(\tilde{r})  + \sin{\tilde{r} \zeta} \ \tilde{\beta}(\tilde{r})\right) \ \mathrm{d} \tilde{r} \ ,
\end{split}
\end{align}
where $ \tilde{\alpha}, \tilde{\beta}$ are piecewise constant functions related to those of the appendix by $\tilde{\alpha}(\tilde{r}) = \alpha\left(r\right)$ and $\tilde{\beta}(\tilde{r}) = \beta\left(r\right)$. As discussed in \ref{appendix:line_to_line}, the particular expressions of $\alpha$ and $\beta$ are given by \eqref{eq:alpha_beta}, where $\alpha_{\Omega_R^+}$, $\beta_{\Omega_R^+}$, and $r_i^+$ are defined in \eqref{eq:alpha}, \eqref{eq:beta}, and \eqref{eq:r_i_plus}, respectively, with $a=D_i, \ b=D_i+H_i, \ a^\prime=D_j, \ b^\prime=D_j+H_j$. Similarly, $\alpha_{\Omega_R^-}$, $\beta_{\Omega_R^-}$, and $r_i^-$ are given by the same expressions but with $a=D_j, \ b=D_j+H_j, \ a^\prime=D_i, \ b^\prime=D_i+H_i$.
While writing down the expressions for $\alpha$ and $\beta$ can be laborious, note that they are relatively simple functions -- they are piecewise constants with up to five pieces, depending on the parameters.

By inspecting \eqref{eq:L2L_r}, it is clear that the term proportional to $\tilde{\alpha}$ is of the same type as the integral treated in the last Section \ref{section:line_to_point}, hence we use the method described there to compute.
As for the term proportional to $\tilde{\beta}$, it can be integrated analytically:
\begin{align}
\label{eq:sin_analytical}
\begin{split}
    \int_{\tilde{r}_1}^{\min\{\tilde{r}_4, \ \tilde{r}_h\}}  \sin{\tilde{r} \zeta} \ \tilde{\beta}(\tilde{r}) \ \mathrm{d} \tilde{r}  = 
    \frac{1}{\zeta} \sum_{i \in \{+, -\}} & \Big[  \cos  \left( \zeta \ \tilde{r}_1^i \right)   -\cos  \left(\zeta \min \{\tilde{r}_2^i, \ \tilde{r}_h \} \right) \\
    &-  \cos  \left( \zeta \min \{\tilde{r}_3^i, \ \tilde{r}_h \}\right)+ \cos  \left( \zeta \min \{\tilde{r}_4^i, \ \tilde{r}_h \}  \right)\Big] \ .
\end{split}
\end{align}

Finally, we compute \eqref{eq:L2L_r} by using \eqref{eq:l2p_final} for each piece of $\tilde{\alpha}$ and \eqref{eq:sin_analytical}.

\subsection{$\zeta$ discretization}
\label{section:discretization}

The next step is to treat the $\zeta$ integral numerically. Since $v_{i \rightarrow j}$ is oscillatory in our cases of interest, integrating this directly with quadratures could be problematic. However, thanks to the separation into blocks, with the exception of the first block $k=1$, the integrand is proportional to $e^{-\zeta^2 N_{k-1} \Delta \tilde{t}}$, as seen in \eqref{eq:source_block}. This has two effects. First, the amplitude of the oscillations is dramatically reduced as $\zeta$ grows. Second, we can cut off the infinite integral up to a certain finite value $b_k$. This suppressing effect is enough to make (adaptive) Gaussian quadrature viable again to perform the integration.
We dedicate Appendix \ref{appendix:cutoff} to show how to choose $b_k$ in each case in a way such that the error in the integral is below a tolerance $\epsilon$.

Unlike the rest of the integrand in \eqref{eq:block_integral}, the source function $s_k$ evolves over time. However, we want to fix the nodes and weights of the quadrature for each pair $i, j$ before the simulation for computational efficiency. This can be done by applying the adaptive Gauss-Kronrod quadrature \cite{kronrod} to an approximation of each of the integrands in Equation \eqref{eq:block_integral}:
\begin{align}
\label{eq:gk_approx}
    I^{i, j}_k(\zeta) = Q\left( N_k - N_{k-1} \right) \left( e^{-\zeta^2 N_{k-1} \Delta \tilde{t}} -  e^{-\zeta^2 N_{k} \Delta \tilde{t}} \right) \frac{v_{i\rightarrow j}(\zeta)}{\zeta^2} \ ,
\end{align}
with $Q=\max_i \ \lvert q(i\Delta \tilde{t})\rvert$.

This works as is in the point to point case \eqref{eq:v_P2P}, but in the line to point \eqref{eq:v_L2P} and line to line \eqref{eq:v_L2L} cases, $v$ has an integral that needs to be performed numerically. It can be costly to execute the adaptive algorithm in this case, because the numerical integral needs to be computed at each evaluation of $I^{i, j}_k(\zeta)$. An alternative approach is to approximate $v$ with elementary functions. A simple way to do this is to apply the trapezoidal rule to the integral in $v$. In the line to point case, and according to Equation \eqref{eq:L2P_integral_cosh}, we can approximate $v$ by
\begin{align}
\label{eq:v_approx}
    \frac{v^\text{L2P}_{i\rightarrow j}(\zeta)}{\zeta^2} \propto \frac{1}{\zeta}\int_{\text{arccosh }\frac{a}{\sigma} }^{\text{arccosh }\frac{b}{\sigma} }  \sin \left( \zeta \ \tilde{\sigma} \cosh{y} \right) \mathrm{d}y \simeq  \frac{1}{2\zeta}\left( \text{arccosh }\frac{b}{\sigma} -\text{arccosh }\frac{a}{\sigma}  \right)\left( \sin{\zeta \frac{b}{r_b}} - \sin{\zeta \frac{a}{r_b}} \right) \ .
\end{align}
While this is a rough estimation, it does at good job at emulating the original function. Critically, the approximation \eqref{eq:v_approx} oscillates at two different frequencies, which is enough for the adaptive algorithm to set integration nodes in a suitable fashion, even if the actual integrand has a slightly different phase or amplitude.

In the line to line case, the relevant integral is \eqref{eq:L2L_r}. As one can see, the integral has two terms: one exactly like the line to point case, and another that can be integrated analytically. Therefore, in this case we can still use \eqref{eq:v_approx} together with the analytical result \eqref{eq:sin_analytical}.
 
By applying the described procedure, the Gauss-Kronrod algorithm will produce $N_{ij,k}$ nodes $\zeta_s$ and weights $w_s$ so that
\begin{align*}
    \int_0^\infty  \left( 1 - e^{-\zeta^2 \Delta \tilde{t}} \right) \frac{v_{i\rightarrow j}(\zeta)}{\zeta^2}  s^i_k \left(\zeta, n \Delta \tilde{t} \right) \ \mathrm{d} \zeta =  \sum_{s=1}^{N_{ij,k}} w_s \left( 1 - e^{-\zeta_s^2 \Delta \tilde{t}} \right) \frac{v_{i\rightarrow j}(\zeta_s)}{\zeta_s^2}  s^i_k \left(\zeta_s, n \Delta \tilde{t} \right) \ ,
\end{align*}
for $k > 1$.
Therefore, we need to evaluate the function $v_{i\rightarrow j}$ corresponding to each case at the particular nodes $\zeta_s$ selected by the adaptive algorithm.

As a remark, note that the quantities
\begin{align}
\label{eq:precomputation}
    H^{i \rightarrow j}_{k, s} = \frac{1}{k_g} \frac{\alpha}{r_b^2} \, w_s \left( 1 - e^{-\zeta_s^2 \Delta \tilde{t}} \right) \frac{v_{i\rightarrow j}(\zeta_s)}{\zeta_s^2} \ ,
\end{align}
only need to be computed once for each simulation, and that the temperature difference in terms of $H$ is written as
\begin{align}
\label{eq:single_interaction}
    T_{i\rightarrow j}(n \Delta \tilde{t}) - T_0  = \sum_{k= K_{i \rightarrow j}(\epsilon)}^K \sum_{s=1}^{N_{ij,k}} H^{i \rightarrow j}_{k, s} \cdot s^i_k \left(\zeta_s, n \Delta \tilde{t} \right) \ .
\end{align}
We will refer to $H$ as the \textit{precomputation} of this method because of this reason.

\subsection{Error tolerance}
Until now, we have written all expressions with a generic error tolerance $\epsilon$. In order for our final result to have a precision below the tolerance that we provide, we must adjust the tolerances of each formula accordingly.

In our case, we want to compute \eqref{eq:block_integral} with an error tolerance of $\epsilon$. First, note that \eqref{eq:block_integral} contains a sum over the blocks. Therefore, each term inside them sum must yield an error of $\epsilon/K$ at most. This is the tolerance that we need to use in \eqref{eq:N_r_eps}, and also in the adaptive algorithm to compute the nodes for the integrand \eqref{eq:gk_approx} (which approximates the integrand in \eqref{eq:block_integral}). This includes the cutoffs of the $\zeta$ integral described in Appendix \ref{appendix:cutoff}.

The last tolerance left to discuss is that of the computation of $\frac{v(\zeta)}{\zeta^2}$, detailed in Section \ref{section:evaluation}. The choice of tolerance affects the value of $h$ in \eqref{eq:v_L2P_bar} and subsequent, for the line to point, and in \eqref{eq:L2L_sin} and subsequent, for the line to line.
Let us denote this tolerance by $\epsilon^\prime$.
According to the discussion in Appendix \ref{appendix:cutoff}, we see that it should be roughly satisfied that
\begin{align*}
    \frac{\epsilon}{K} \simeq \frac{Q}{4 \pi^2 k_g} \epsilon^\prime \ln{\frac{N_k-1}{N_{k-1}}} \ .
\end{align*}
This gives a criterion to choose $\epsilon^\prime$.

\subsection{The first block}
\label{section:first_block}
The first block $s_1$ is only relevant in close interactions, for example self-responses. However, the procedure described above will not work well in this case. This is because the assumption that the oscillating integrals in Equation \eqref{eq:block_integral} can be performed using standard quadrature methods does not hold for $k=1$. Indeed, $s_k$ is proportional to $e^{-\zeta^2 N_{k-1} \Delta \tilde{t}}$, but since $N_0 = 0$, this property is tautological for $s_1$. 
Therefore, the direct use of Gauss quadratures becomes problematic and requires many integration points to obtain good results.

An alternative is to use the specialized Bakhalov-Vasil'eva method for $k=1$, as it was used in \cite{non-history}. We outline the method in Appendix \ref{appendix:bakhalov}.

\section{An illustrated example}
\label{section:example}

As we have discussed in Sections \ref{section:methodology} and \ref{section:evaluation}, choosing blocks is equivalent to partitioning the timeframe of the problem, which effectively separates the different time scales that exist in it. Also, each block is associated with a region of space where it has influence, effectively separating space scales.
These two mechanisms are implemented in the present method in the form of domain reduction and gaussian suppression, as shown in Figure \ref{fig:line_to_point_block}.

\begin{figure}[h!]
    \centering
    \includegraphics[scale=0.8]{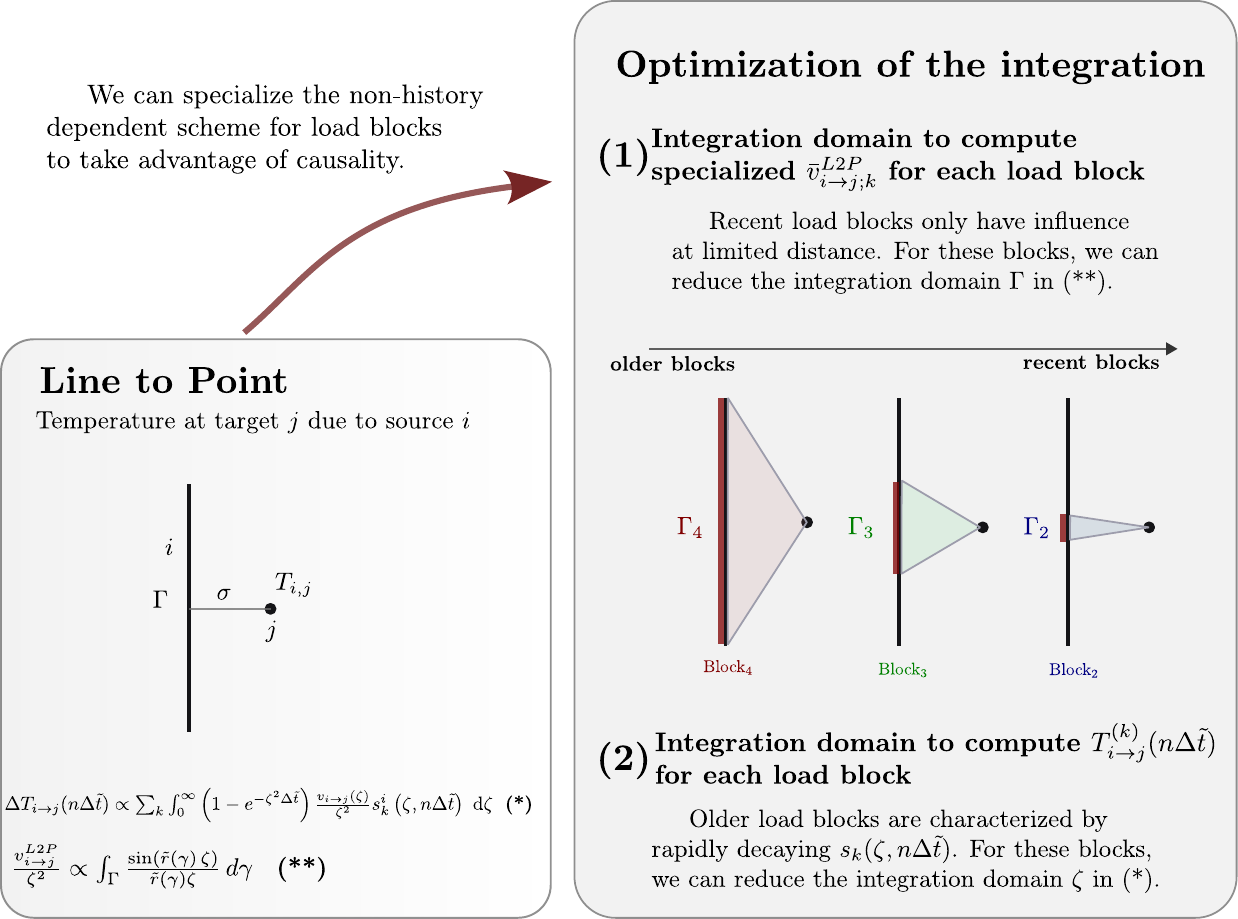}
    \caption{Summary of the integration strategy for the line to point case. The blocks structure introduces two different, complementary factors that help reduce the numerical complexity of the integral:
    \ \ (\textit{i}) Within a given block $s_k$, interactions have bounded time to occur. We reduce the effective line integration region by removing those that do not have time to influence the target. This is specially relevant in the newer blocks.
    \ \ (\textit{ii}) Block $s_k$ is proportional to $e^{-\zeta \Delta \tilde{t} N_{k-1}}$, which reduces the effective integration region over $\zeta$. This is specially relevant in the older blocks.
    } 
    \label{fig:line_to_point_block}
\end{figure}

We explore these two effects in detail by examining two concrete examples, both featuring the line to point case, with $D=0 \text{ m}$, $H=150 \text{ m}$, and $z=75 \text{ m}$, at two distinct values of $\sigma$ -- a short one and a long one. The purpose of these setups is that in each of them, one of the mechanisms dominates over the other so they become clearly identifiable. Of course, in intermediate ranges, both mechanisms play a role in the better behaviour of the integrand.
We apply a unit step as the load at the source.
The calculations of the resulting temperature responses assume a thermal diffusivity $\alpha = \,$ \SI{1e-6}{\square\meter\per\second}, a thermal conductivity $k = \,$ \SI{3}{\watt\per\meter\per\kelvin}, a borehole radius $r_b = \,$ \SI{0.1}{\meter}, and a time step $\Delta t = \,$ \SI{1}{\hour}.

\subsection{Short distances: reduced source region}

Let us set $\sigma = 5 \text{ m}$. In this setup, the minimum distance $\sigma$ from the line source to the target point is much smaller than the length of the line $H$. This allows to illustrate the reduction of the original line source to a smaller effective one, as explained in Section \ref{section:blocks}. 
The effect of this reduction on the integrand of \eqref{eq:block_integral} is shown in Figures \ref{fig:short_block_2}, \ref{fig:short_block_3}, for the second and third blocks, respectively. In particular, it can be seen that the reduced versions of the integrand are much smoother than the full versions, which leads to fewer Gaussian quadrature points.

We also show the integrand for the first block in Figure \ref{fig:short_block_1}. As shown, the integrand is highly oscillating and is a bad candidate for Gaussian quadratures. This illustrates the reason why the first block needs to be treated separately, as discussed in Section \ref{section:first_block}

\begin{figure}[htbp]
    \centering
    \includegraphics[width=0.6\textwidth]{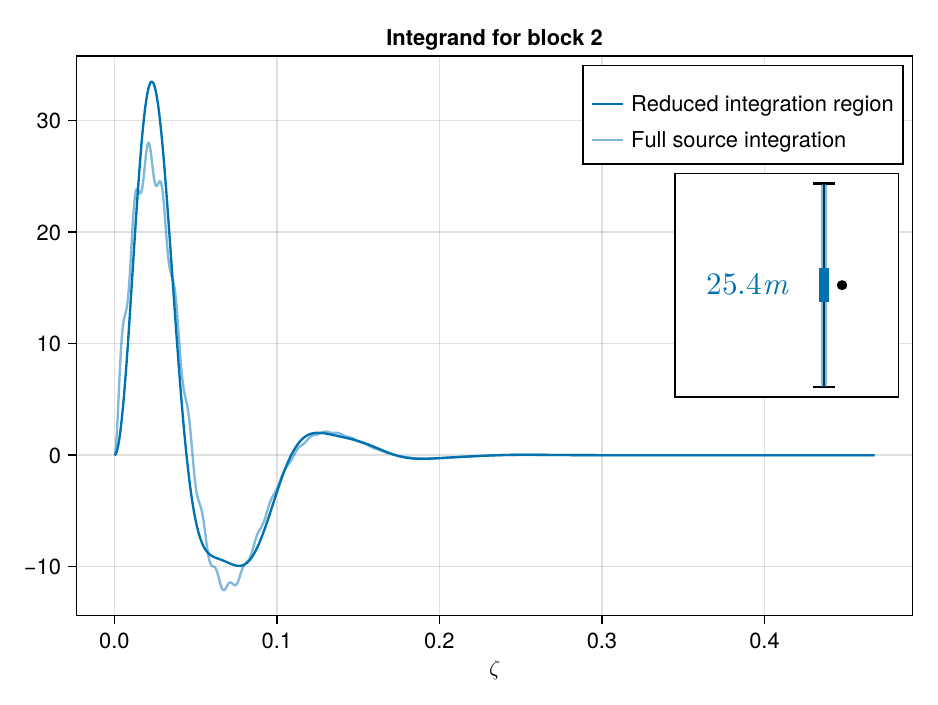}
    \caption{Integrand in equation \eqref{eq:block_integral}, for the second block, in two cases: (\textit{i}) using the reduced effective source in dark blue, and (\textit{ii}) using the full source, in light blue. }
    \label{fig:short_block_2}
\end{figure}
\begin{figure}[htbp]
    \centering
    \includegraphics[width=0.6\textwidth]{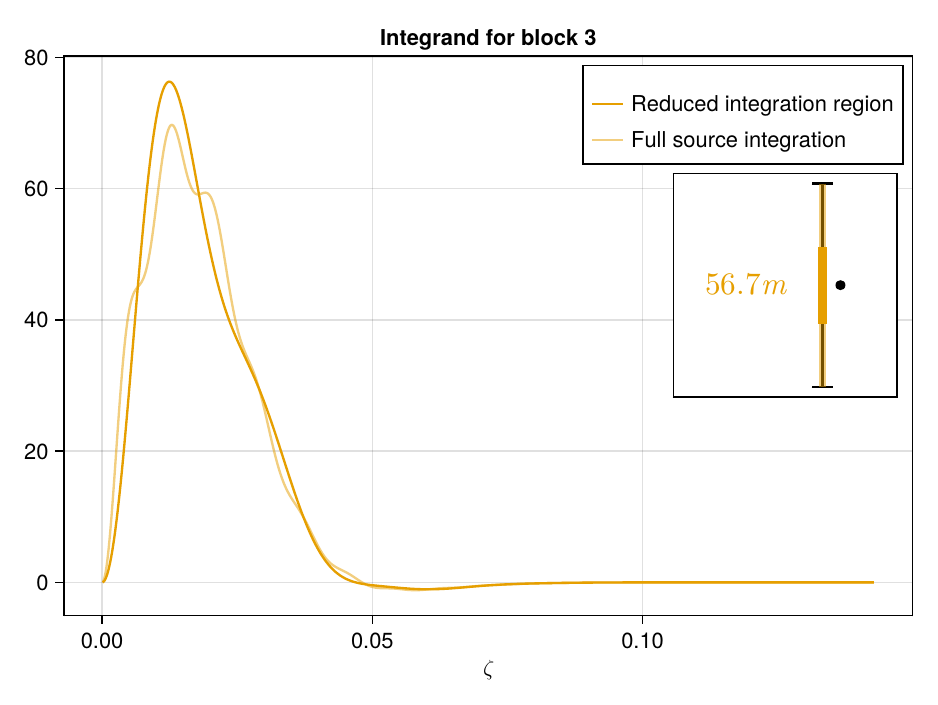}
    \caption{Integrand in equation \eqref{eq:block_integral}, for the third block, in two cases: (\textit{i}) using the reduced effective source in dark yellow, and (\textit{ii}) using the full source, in light yellow. }
    \label{fig:short_block_3}
\end{figure}
\begin{figure}[htbp]
    \centering
    \includegraphics[width=0.6\textwidth]{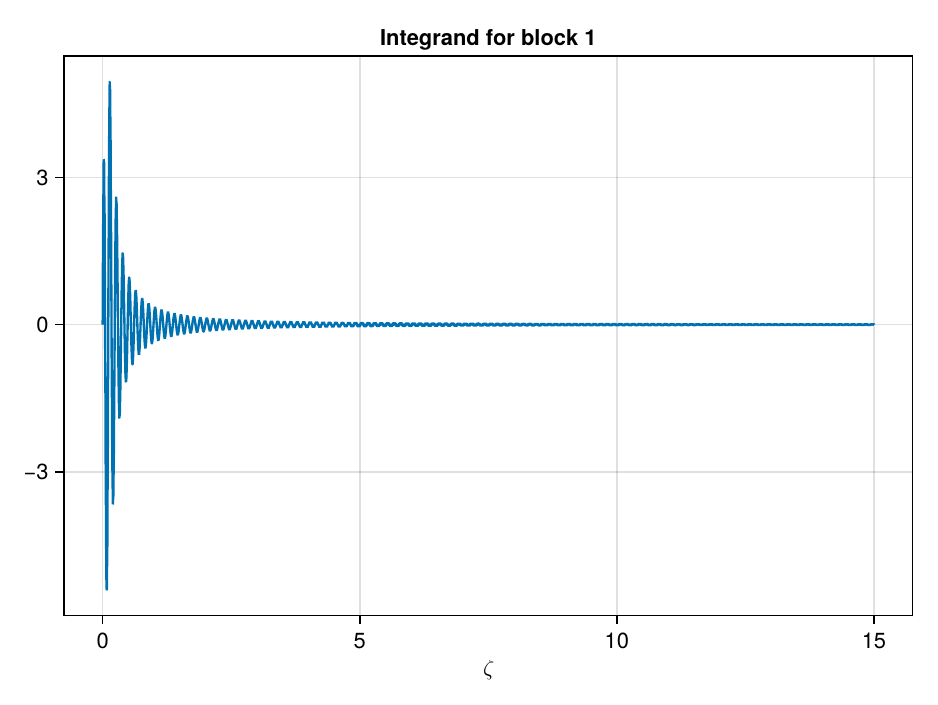}
    \caption{Integrand in equation \eqref{eq:block_integral} for the first block.}
    \label{fig:short_block_1}
\end{figure}
\clearpage

\subsection{Long distances: gaussian suppression}

Let us set $\sigma = 50m$. In this setup, the minimum distance $\sigma$ from the line source to the target point is large, so that it takes a long time for the causal influence of the source to reach the target, therefore $N_1$ is large. This implies that the second block, proportional to $e^{-\zeta N_1\Delta\tilde{t}}$, 
introduces a suppressing gaussian factor in the integrand of \eqref{eq:block_integral}.
The effect of this factor is shown in Figures \ref{fig:long_block_2}, \ref{fig:long_block_3}, and it can be summarized in two things: the effective upper limit of the $\zeta$ integral is reduced, while the amplitude of the oscillations is vastly decayed. Both effects contribute to the lesser need of integration quadrature points.

\begin{figure}[htbp]
    \centering
    \includegraphics[width=0.6\textwidth]{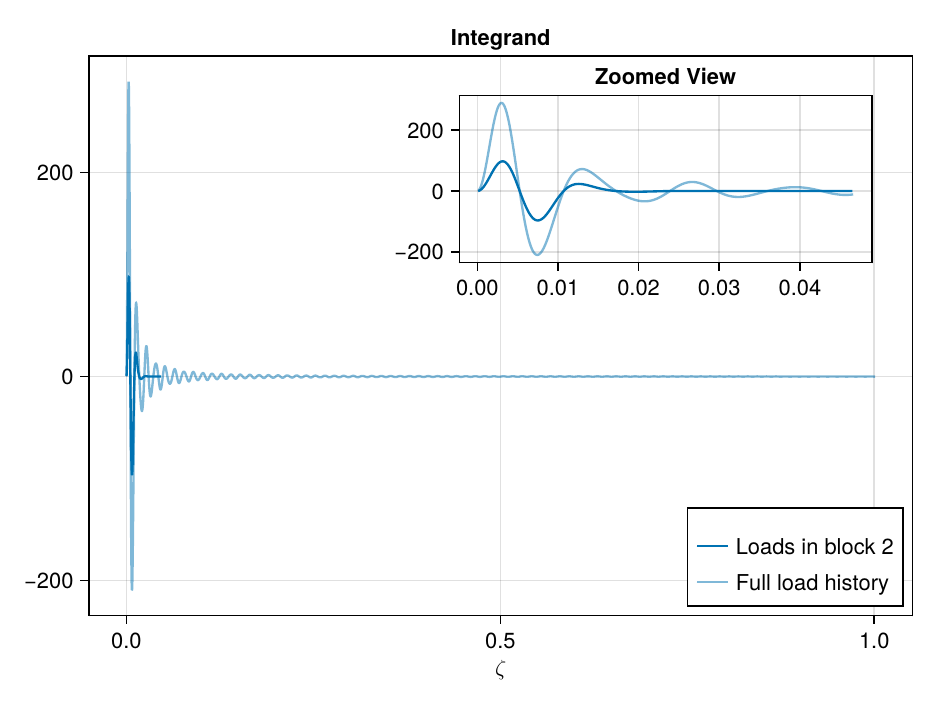}
    \caption{Integrand in equation \eqref{eq:block_integral} in two cases: (\textit{i}) only considering the second block, in dark blue, and (\textit{ii}) using the full source function, in light blue. }
    \label{fig:long_block_2}
\end{figure}
\begin{figure}[htbp]
    \centering
    \includegraphics[width=0.6\textwidth]{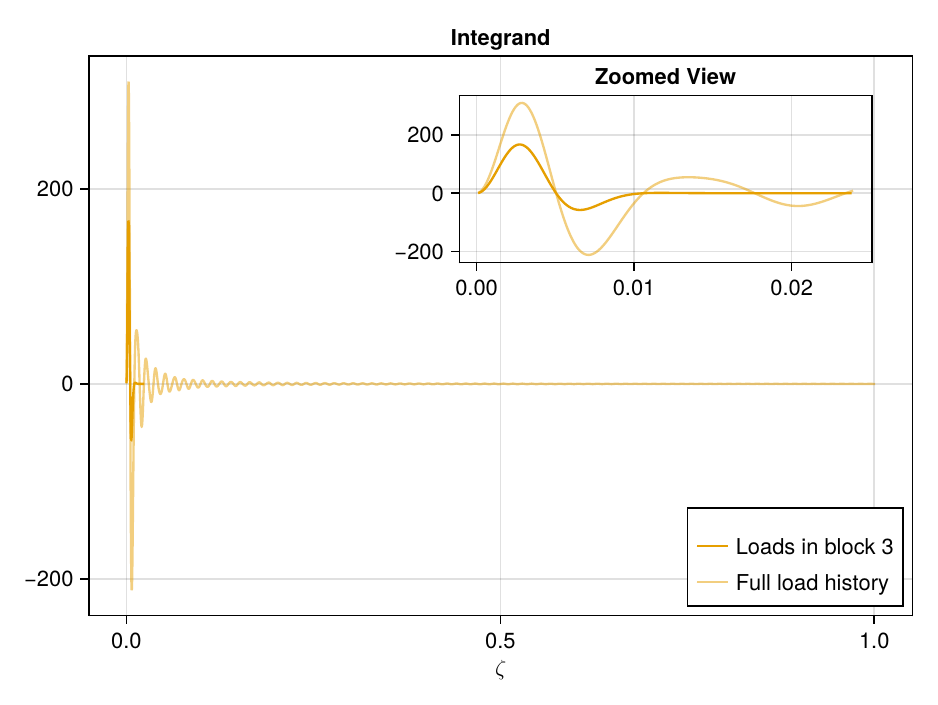}
    \caption{Integrand in equation \eqref{eq:block_integral} in two cases: (\textit{i}) only considering the third block, in dark yellow, and (\textit{ii}) using the full source function, in light yellow. }
    \label{fig:long_block_3}
\end{figure}

\section{Result analysis}
\label{section:results}

We assess the proposed method in two aspects: computational efficiency and accuracy.
In order to do so, we will look at the example of one source and one target at a distance enough so that the block $s_1$ can be disregarded. Note that since the computation of $s_1$ can be done with the method proposed in \cite{non-history} that have already been validated there, we only assess blocks $s_k$ for $k>1$.

\subsection{Error analysis}

In order to measure the error, we compare the result with the discrete convolution of the load with the impulse response. In practice, we evaluate the equivalent formula
\begin{align}
\label{eq: convolution}
    T(t) - T_0 = \sum_{i=0}^{N_t} \Big( q^\prime \big((N_t - i - 1) \Delta t \big) - q^\prime \big((N_t - i ) \Delta t\big) \Big) h\left(i \Delta t\right) \ ,
\end{align} 
using the step response $h$ of the system.
The step response of the point source is given by 
\begin{align}
\label{eq:real_point_to_point}
    h^{P2P}(t, r) = \frac{\text{erfc}\left(\frac{r}{\sqrt{4 t \alpha}}\right)}{4\pi r k_g} \ , 
\end{align}
and the responses of a line source evaluated at a point and its mean along another line are, respectively:
\begin{align}
\label{eq:real_line_to_point}
    h^{L2P}(t, \sigma, z) &= \int_{D}^{D+H} h^{P2P}\left(t, \sqrt{\sigma^2 + (z-z^\prime)^2}\right) \ \mathrm{d}z^\prime \ , \\
\label{eq:real_line_to_line}
    h^{L2L}(t, \sigma) &= \frac{1}{H_t}\int_{D_t}^{D_t+H_t} \int_{D_s}^{D_s+H_s} h^{P2P}\left(t, \sqrt{\sigma^2 + (z_s-z_t)^2}\right) \ \mathrm{d}z_s \mathrm{d}z_t \ .
\end{align}
For the purpose of assessing the error of the blocks method in a range of cases, we perform the simulation of $4$ years, with hourly time steps, with several distance values, with two different loads: a constant, unit step load, and the synthetic load 
\begin{align}
\label{eq:load}
    q^\prime(t) = 20 \sin{\left(\frac{2 \pi t}{8760}\right)} + 5\sin{\left(\frac{2 \pi t}{24}\right)} + 5 \ ,
\end{align}
which aims to replicate seasonal and daily variations in load demand.

Figures \ref{fig:point_to_point_error}, \ref{fig:line_to_point_error}, and \ref{fig:line_to_line_error} show $\Vert \varepsilon \Vert_\infty$, the maximum value over time of the absolute error between the two temperatures defined by \eqref{eq:block_integral} and \eqref{eq: convolution} for different tolerance $\epsilon$ and distance values, for the point to point, line to point, and line to line cases, respectively.
Note that we used double accuracy precision ($\varepsilon \simeq 2 \cdot 10^{-16}$) to compute the integrals \eqref{eq:real_line_to_point} and \eqref{eq:real_line_to_line} used in the reference temperature \eqref{eq: convolution}.

\begin{figure}[h!]
    \centering
    \includegraphics{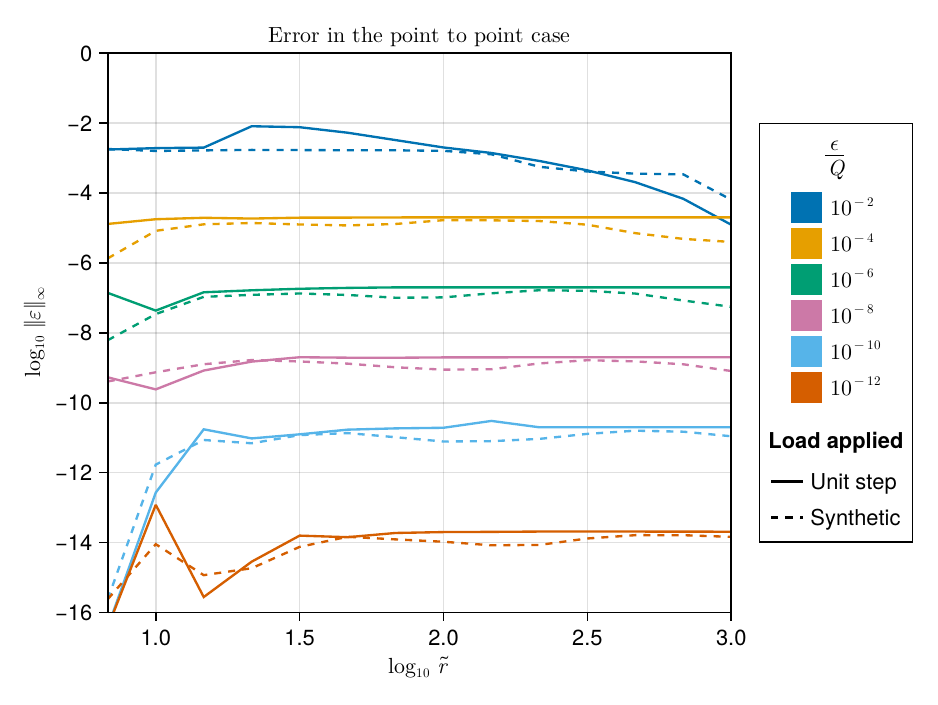}
    \caption{Logarithm of the $\ell^\infty$ norm of the error of the blocks method, for several error tolerances. The computation has been made with two different loads: (\textit{i}) with a constant unit step in solid lines, and (\textit{ii}) with the synthetic load \eqref{eq:load} in dashed lines.
    }
    \label{fig:point_to_point_error}
\end{figure}
\begin{figure}[h!]
    \centering
    \includegraphics{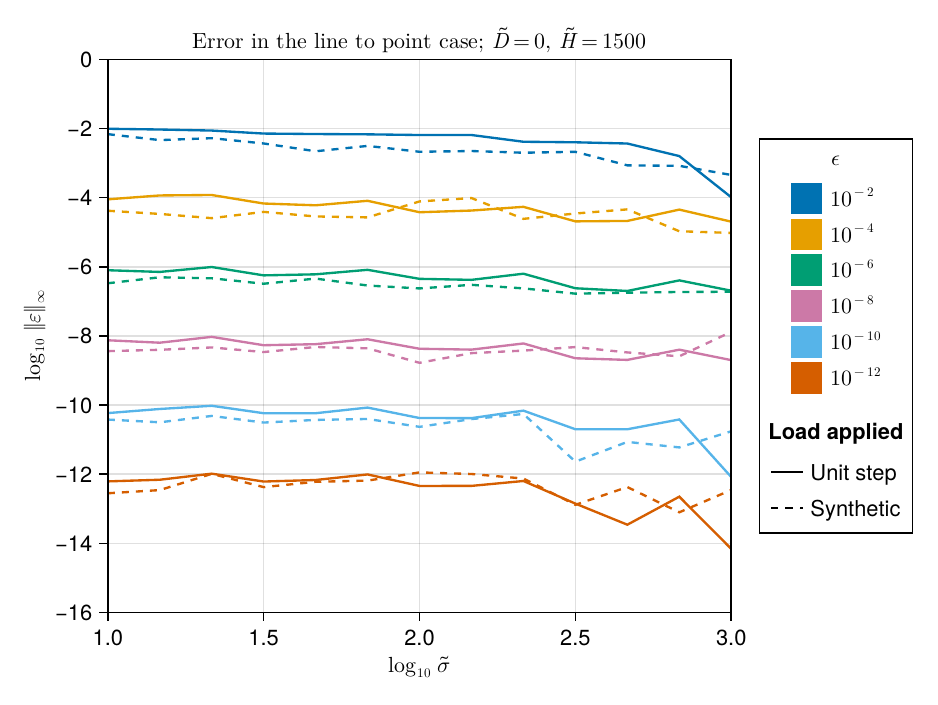}
    \caption{L2P}
    \label{fig:line_to_point_error}
\end{figure}
\begin{figure}[h!]
    \centering
    \includegraphics{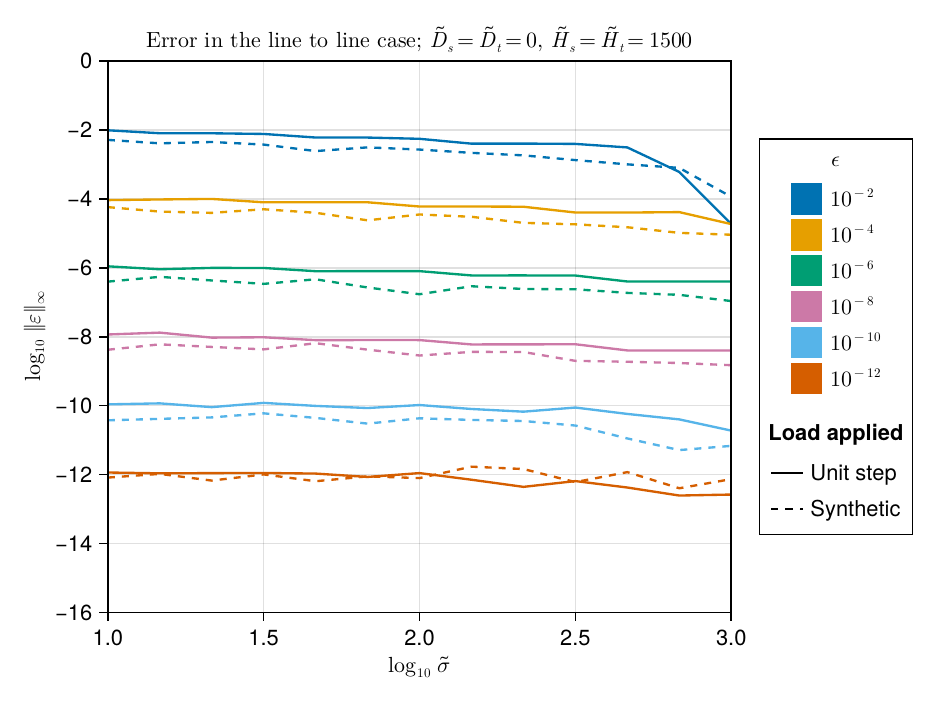}
    \caption{L2L}
    \label{fig:line_to_line_error}
\end{figure}

The results show that, for all tolerances $\epsilon$ tested, $\Vert \varepsilon \Vert_\infty < \epsilon$ is kept below the specified tolerance. Moreover, this happens not by a large margin, so we are not incurring in more additional computations than needed to satisfy the error tolerance. 

\subsection{Computational complexity}

We now examine how costly it is to use the method presented in Section \ref{section:evaluation}. For a single interaction, we need to compute \eqref{eq:single_interaction}. Focusing on the precomputation first, we see that one needs to compute $H^{i\rightarrow j}_{k, s}$ defined in \eqref{eq:precomputation} for each point in the $\zeta$ discretization, of which there are $N_\zeta \vcentcolon = \sum^K_{k=K_{i\rightarrow j} (\epsilon)} N_{ij, k}$. Now, the complexity of computing each $H^{i\rightarrow j}_{k, s}$ depends on the specific form of $v_{i\rightarrow j}$ in each geometrical configuration.

\begin{itemize}
    \item Point to point (Section \ref{section:point_to_point}). A single function evaluation is needed, hence the compelxity is constant $\mathcal{O}(1)$.
    \item Line to point (Section \ref{section:line_to_point}). This case is slightly more involved. The integral is divided in subsegments according to the value of the discontinuous function \eqref{eq:ltp_alpha}, as written in \eqref{eq:L2P_integral_cosh}. We use two distinct methods to perform this integral: adaptive Gaussian integration and the asymptotic method.
    For the asymptotic method, assume that we use truncate at $n$ terms the asymptotic expansion. We need to compute the coefficients \eqref{eq:sigma_functions_cosh_a} that are used in \eqref{eq:asymptotic_formula} to compute the integrals.
    The former has complexity $\mathcal{O}(n^2)$ but only occurs once, and the latter also has complexity $\mathcal{O}(n^2)$ but can occur multiple times per integral evaluation. However, $n$ is usually taken to be small. Since the coefficients \eqref{eq:sigma_functions_cosh_a} are defined recursively by multiplying increasing integers at each iteration, they become numerically unstable for $n$ too big. A good balance between stability and accuracy is around $n=10$.
    
    The number of operations performed in the Gaussian integration is harder to compute exactly, since the adaptive algorithm makes a variable number of them. 
    \item Line to point (Section \ref{section:line_to_line}). According to the discussion, the complexity of this case is comparable to that of the line to point case.
\end{itemize}

\subsection{Comparison with original method}

Although it is difficult to directly compare the present method to the original one by only looking at the number of operations, we can easily compare the number of discretization points $N_D$ used in each of them. 

For the blocks method this is simple $N_D = N_\zeta$ for all cases.
However, for the original formulation, since we must take into account both the $\zeta$ discretization points $N_\zeta$ and the line discretization points $N_l$, $N_D$ has a different expression in each case. We have:
\begin{itemize}
    \item Point to point: $N_D = N_\zeta$.
    \item Line to point: $N_D = N_\zeta N_l$.
    \item Line to line: $N_D = N_\zeta N_l^2$.
\end{itemize}

\begin{figure}[h!]
    \centering
    \includegraphics{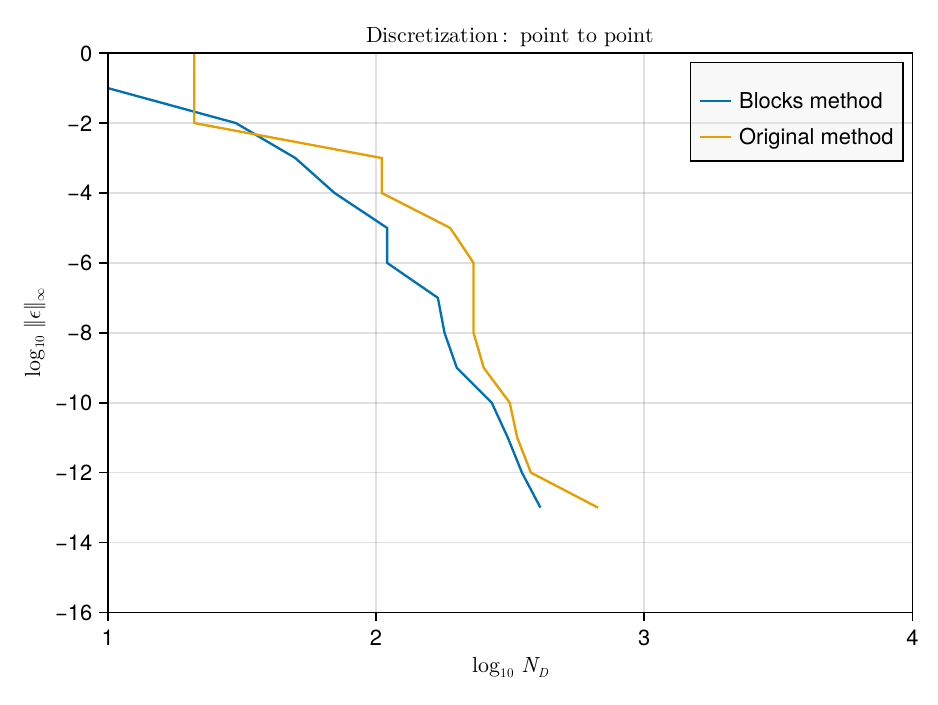}
    \caption{Logarithm of the error $\epsilon$ of the blocks method (with respect to the convolution \eqref{eq: convolution}) against the number of discretization points $N_D$, for both the blocks and the original method, in the point to point case. The geometric parameters used were $r=1\text{m}$.}
    \label{fig:point_to_point_disc}
\end{figure}
\begin{figure}[h!]
    \centering
    \includegraphics{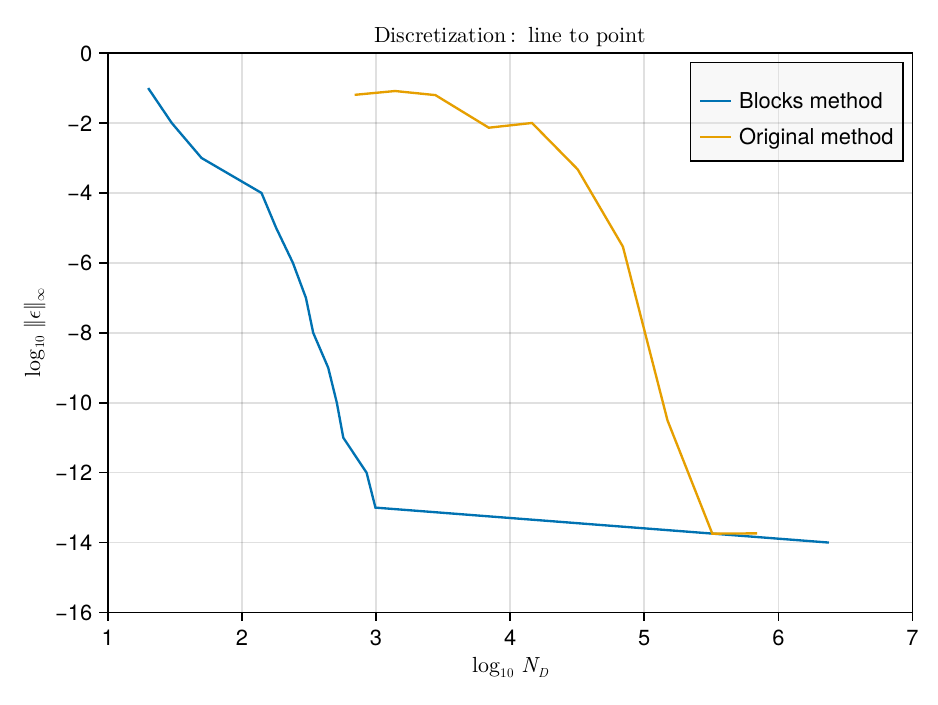}
    \caption{Logarithm of the error $\epsilon$ of the blocks method (with respect to the convolution \eqref{eq: convolution}) against the number of discretization points $N_D$, for both the blocks and the original method, in the line to point case. The geometric parameters used were $D=0, \ H=150\text{m}, \ z=75\text{m}, \  \sigma=1\text{m}$.}
    \label{fig:line_to_point_disc}
\end{figure}

\begin{figure}[h!]
    \centering
    \includegraphics{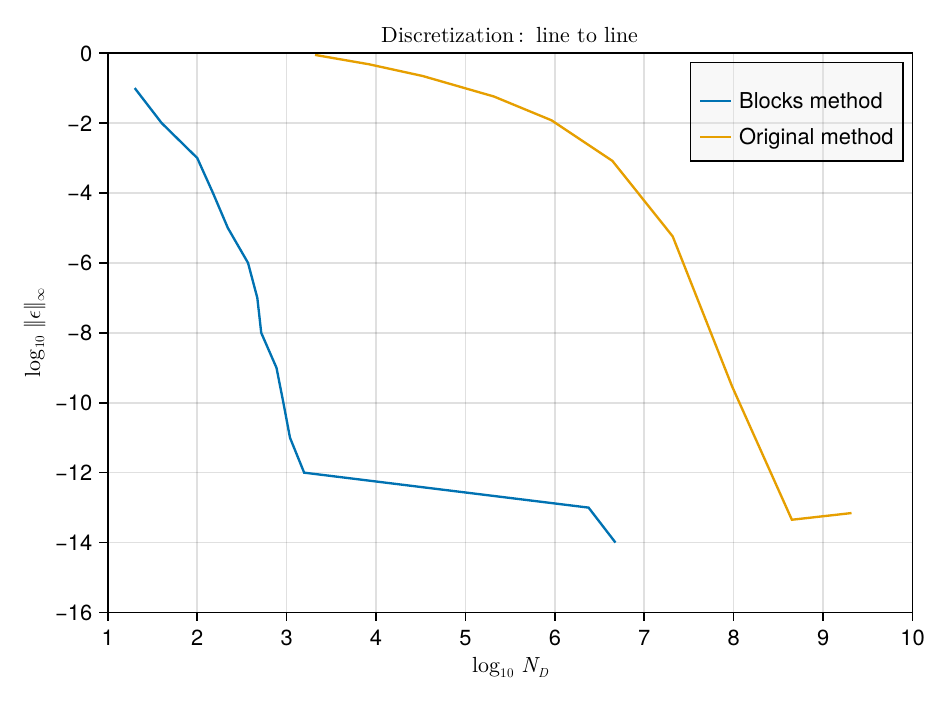}
    \caption{Logarithm of the error $\epsilon$ of the blocks method (with respect to the convolution \eqref{eq: convolution}) against the number of discretization points $N_D$, for both the blocks and the original method, in the line to line case. The geometric parameters used were $\tilde{D}=0, \ \tilde{H}=1500, \ \tilde{D}^\prime=0, \ \tilde{H}^\prime=1500, \  \tilde{\sigma}=10$.}    \label{fig:line_to_line_disc}
\end{figure}

The comparison between $N_D$ in each case for both methods is shown in Figures \ref{fig:line_to_point_disc} and \ref{fig:line_to_line_disc}.
Recall that the original method cannot compute the solution up to a prescribed error tolerance, while the present method has been designed with this feature in mind.
Therefore, for the ``blocks method" curve, we use as an input the error tolerance and we get as an output the number of points $N_D$, while for the ``original method" curve, we use as an input the number of points and get as an output the error.

In Figure \ref{fig:point_to_point_disc}, we can see that the present method needs approximately the same amount of points for a given accuracy than the original one.
In contrast, it is clear from Figures \ref{fig:line_to_point_disc} and \ref{fig:line_to_line_disc} that the present method, for both line source cases, needs orders of magnitude less points $N_D$ than the original one to achieve a given accuracy. In the line to point case, this is roughly 2 orders of magnitude of difference while in the line to line case, it is from 4 to 5.
An important conclusion from the results is that from roughly $\epsilon_t = 10^{-12}$, the required number of needed discretization points $N_D$ suddenly experiences a rapid increase. The increase is due to the adaptive algorithm for the $\zeta$ integral applied to the function \eqref{eq:v_approx}. This can be understood by looking at the examples in Section \ref{section:example}. As the tolerance decreases, the effect of the two mechanisms illustrated in Figures \ref{fig:short_block_2}, \ref{fig:short_block_3}, \ref{fig:long_block_2}, \ref{fig:long_block_3} is less pronounced, making it resemble
more the full integrand, which makes \eqref{eq:v_approx} more difficult to integrate  up to the prescribed accuracy with Gaussian quadratures, resulting in a large amount of integration nodes.
While this does not mean the tolerance can not be met below $\epsilon_t$, it means that the speed and utility of the method are severely reduced for such accuracies. However, $\epsilon_t$ is sufficiently low that the method exhibits a big speedup with respect to the original one for most practical purposes.

Finally, we discuss execution time. It is our opinion that this is not the best way of comparing algorithms. This is because the execution time can depend on many factors, from the hardware, through the memory state of the computer, to the fine details of the implementation of the algorithm.
However, we realize that in practice, one of the most relevant metrics is execution time.
Therefore, we think it is also important to show the time scales involved in the execution of the methods.
For this case, we compare the original non-history, the blocks, and the convolution (using the FFT algorithm) methods.
We are including the convolution in this comparison for reference and the sake of completeness. However, note that while the non-history based methods are time marching scheme, the convolution is not, which means they are not directly comparable. This is because the non-history approach is applicable to a larger set of problems, for example, it allows to simulate scenarios where the source intensities are not known ahead of time.
The results are shown in Table \ref{table:precomputation} and Table  \ref{table:execution}.
Both tables have been obtained by measuring executions times on the same machine, and optimizing both non-history implementations using the same techniques, to the best of our abilities.

\begin{table}[h!]
\begin{center}
\begin{tabular}{||c || c | c | c | c | c | c||} 
 \hline
 Precomputation method & $\epsilon=10^{-2}$ & $\epsilon=10^{-4}$ & $\epsilon=10^{-6}$  & $\epsilon=10^{-8}$ & $\epsilon=10^{-10}$ & $\epsilon=10^{-12}$ \\ [0.5ex] 
 \hline\hline
 Original (P2P) & 53 $\mu$s  & 313 $\mu$s & 575 $\mu$s & 575 $\mu$s & 782 $\mu$s & 911 $\mu$s \\ 
 \hline
 Blocks (P2P)   & 125 $\mu$s & 138 $\mu$s & 142 $\mu$s & 155 $\mu$s & 166 $\mu$s & 174 $\mu$s \\
 \hline
Convolution (P2P)  & 1.4 ms & 1.4 ms & 1.4 ms & 1.4 ms & 1.4 ms & 1.4 ms\\
\hline\hline
Original (L2P) & 592 $\mu$s & 14 ms & 65 ms & 80 ms & 112 ms & 566 ms \\
\hline
Blocks (L2P) & 346 $\mu$s & 772 $\mu$s & 1 ms & 2 ms & 4 ms & 10 ms \\
 \hline
Convolution (L2P)  & 190 ms & 397 ms & 618 ms & 862 ms & 1.2 s & 1.6 s\\
\hline\hline
Original (L2L) & 91 ms & 6.6 s & 74 s & 91 s & 385 s & 689 s  \\
\hline
Blocks (L2L) & 781 $\mu$s & 2 ms & 4 ms & 8 ms & 18 ms & 57 ms \\
\hline
Convolution (L2L)  & 660 ms & 2.0 s & 3.4 s &  5.0 s & 5.5 s & 5.8 s\\
\hline
\end{tabular}
\caption{Execution times of the precomputation for a single interaction between a source and a target, in the original, the blocks, and convolution methods, for several values of the tolerance $\epsilon$ and each of the considered cases: point to point, line to point, and line to line.}
\label{table:precomputation}
\end{center}
\end{table}

\begin{table}[h!]
\begin{center}
\begin{tabular}{|| c || c | c | c | c | c | c ||} 
 \hline
 Simulation method & $\epsilon=10^{-2}$ & $\epsilon=10^{-4}$ & $\epsilon=10^{-6}$  & $\epsilon=10^{-8}$ & $\epsilon=10^{-10}$ & $\epsilon=10^{-12}$ \\ [0.5ex] 
 \hline\hline
 Original (P2P) & 5 ms & 12 ms & 24 ms & 24 ms & 36 ms & 41 ms \\ 
 \hline
 Blocks (P2P)  & 18 ms & 27 ms & 32 ms & 40 ms & 49 ms & 64 ms\\
 \hline
 Convolution (P2P)  & 4 ms & 4 ms & 4 ms & 4 ms & 4 ms & 4 ms\\
\hline\hline
Original (L2P) & 6 ms & 13 ms & 24 ms & 24 ms & 36 ms & 41 ms \\
\hline
Blocks (L2P) & 18 ms & 30 ms & 47 ms & 52 ms & 77 ms & 116 ms \\
\hline
 Convolution (L2P)  & 4 ms & 4 ms & 4 ms & 4 ms & 4 ms & 4 ms\\
\hline\hline
Original (L2L) & 6 ms & 12 ms & 24 ms & 24 ms & 36 ms & 41 ms  \\
\hline
Blocks (L2L) & 22 ms & 31 ms & 57 ms & 91 ms & 121 ms & 197 ms \\
\hline
 Convolution (L2L)  & 4 ms & 4 ms & 4 ms & 4 ms & 4 ms & 4 ms\\
 \hline
\end{tabular}
\caption{Execution times of a simulation of a single interaction between a source and a target of 20 years with hourly timesteps, using the original, the blocks, and the convolution methods, for several values of the tolerance $\epsilon$ and each of the considered cases: point to point, line to point, and line to line. In the convolution rows, the FFT algorithm has been used, and the tolerances refer to the adaptive quadrature tolerance used to compute the step responses.}
\label{table:execution}
\end{center}
\end{table}

\begin{table}[h!]
\begin{center}
\begin{tabular}{|| c || c | c | c | c | c | c ||} 
 \hline
 Integral computation & $\epsilon=10^{-2}$ & $\epsilon=10^{-4}$ & $\epsilon=10^{-6}$  & $\epsilon=10^{-8}$ & $\epsilon=10^{-10}$ & $\epsilon=10^{-12}$ \\ [0.5ex] 
 \hline
 Blocks (P2P)  & 6 ms & 9 ms & 11 ms & 15 ms & 19 ms & 26 ms\\
\hline
Blocks (L2P) & 6 ms & 10 ms & 18 ms & 21 ms & 33 ms & 51 ms \\
\hline
Blocks (L2L) & 8 ms & 12 ms & 23 ms & 39 ms & 53 ms & 105 ms \\
\hline
\end{tabular}
\caption{Execution times of just the temperature computation step \eqref{eq:single_interaction} (without update step) of a single interaction between a source and a target of 20 years with hourly timesteps, using the blocks methods, for several values of the tolerance $\epsilon$ and each of the considered cases: point to point, line to point, and line to line.}
\label{table:temperature_computation}
\end{center}
\end{table}

There are some comments to make on the precomputation results shown in Table \ref{table:precomputation}.
\begin{itemize}
    \item The precomputation in the blocks method is faster than in the original method: 
    \begin{itemize}
        \item Point to point: between 2 and 5 times faster (except for the lowest tolerance).
        \item Line to point: up to 2.5 orders of magnitude faster.
        \item Line to line: between 3 or 4 orders of magnitude faster.
    \end{itemize}
    \item It is also faster than the convolution, roughly between 2 and 3 orders of magnitude in all cases.
    \item The precomputation in the original method was already much faster than the convolution in the point to point and line to point cases, but it was dramatically slower in the line to line case. The blocks method solves this problem.    \item Note that for the benchmark of the convolution method we are computing the step response $h$ \textit{at each time step} to use in \eqref{eq: convolution}. This is in contrast with the common practice of only evaluating it at a few points and interpolating it, which would be much faster. However, this would not provide a fair comparison between methods. This is because it is not possible to control the error the interpolation commits a priori, which is a focus of the present method. 
    \item In simulations of geothermal fields with many sources, where one typically is interested in the line to line case, and where \eqref{eq:precomputation} must be computed per interaction (which scales with the square of the number of sources), using the blocks method has a big impact, turning the computation from unfeasible to possible.
\end{itemize}

The results of the simulation running time reported in Table \ref{table:execution} also warrant some comments:
\begin{itemize}
    \item The simulation time is higher for the blocks method than both the original one and the convolution. This happens because of the extra computational effort of handling the blocks at each time step. In the original method this was not the case (one could think of it as having only 1 block), while in the convolution method, the computation is entirely different.
    \item In the blocks method, the simulation phase performs mainly two operations at each time step: the update of the source functions \eqref{eq:recursive_sk} and the computation of the temperature \eqref{eq:single_interaction}. The original non-history method also performs those steps, but with different expressions. The key difference in the two of them is the complexity with respect to the  number of sources $N_s$.
    In the blocks method:
    \begin{itemize}
        \item The update step \eqref{eq:recursive_sk} must be computed for each source, yielding $\mathcal{O}(N_s)$.
        \item The computation step \eqref{eq:single_interaction} must be done for each \textit{interaction}, which scale with the square of the number of sources $\mathcal{O}(N_s^2)$. 
    \end{itemize}
    In the original method, both steps must be performed \textit{per interaction}, yielding a global $\mathcal{O}(N_s^2)$. Therefore, the overall scaling as $N_s$ increases is more favorable in the blocks method.
    \item Even though the convolution running time is much faster than the non-history for a single time-sequence computation, recall that we are interested in time-stepping schemes that allow us to compute the temperature without knowing all the loads in advance. This means that, in the convolution method, the simulation phase must be run at each time step, using all the previous loads up to the current one.
    This is not the case for the non-history methods, which are already formulated as marching schemes. 
\end{itemize}

\section{Conclusions}
\label{section:conclusions}

We have presented a new non-history based method applied to point and line sources that improves the performance of the precomputation over the original non-history method while preserving the linear scaling in the number of time steps.
The proposed method achieves better performance by reducing the integration regions appropriately, leading to better behaved integrands. The guiding principle on how to do this is the idea of causality -- effects of sources that have not had time to influence the target (up to a tolerance $\epsilon$) are disregarded.

Another advantage of our approach is that we have a natural way of controlling the error of the solution. This is an improvement over the original algorithm, in which the error could not be reliably predicted a priori. Having such error control is desirable, not only because one can consistently obtain accurate simulations, but also because it allows for a trade-off between performance and precision when needed.

The integrands in the relevant regions allowed us to forego the original integration scheme, based on the Bakhalov-Vasil'eva oscillatory integration method, and that required a large number of integration points, making it computationally expensive.
The method proposed in this paper does not suffer from this problem, allowing us to compute the integrals with a combination of the asymptotic method and adaptive Gaussian quadratures, which are fast. 

The result analysis for a single interaction shows that the precomputation execution time of the blocks method is reduced by up to 4 orders of magnitude compared to the original one in the line to line case, which is one of the most relevant processes used in geothermal simulations. This brings the precomputation time of a single interaction down to between tenths of microseconds to tenths of milliseconds, depending on the tolerance and geometry. This represents a big speedup with respect to the original method, making the non-history scheme feasible for real applications.
The blocks method also improves over the convolution in the precomputation phase by 3 to 4 orders of magnitude. 

A trade-off for the improvement in the precomputation is the increase of computational effort in the simulation phase, which is higher than in both the original method and the convolution. However, this is not expected to be an obstacle for the method. First, the combined time of both the precomputation and the simulation is lower in the blocks method. 
Moreover, the difference in execution time is expected to be reduced for more complex simulations, where the method can leverage its ability to organize time and space scales.

\section*{Acknowledgments}
The authors acknowledge Energimyndigheten (Swedish Energy Agency) for financing this research via the grants P2022-00486, P2022-00499 (TERMO program) and P2022-01040 (JPP SES and GEOTHERMICA ERA-Net network programs).

\begin{appendices}

\section{The Bakhalov-Vasil'eva oscillatory integration method}
\label{appendix:bakhalov}

The Bakhalov-Vasil'eva method \cite{bakhvalov} is a specialized to compute oscillatory integrals of the form
\begin{align*}
    I_f(\omega) = \int_{-1}^{1} f(x)e^{i \omega x} \mathrm{d}x \ .
\end{align*}
The idea of the method is to approximate $f$ as a truncated series expansion of Legendre polynomials of order $n$, and then use their analytical integrals to provide a closed-form approximation for the original integral.
By means of a linear transformation, the method can be applied to arbitrary intervals, yielding
\begin{align*}
    \int_a^b f(x) e^{i \omega x} \ \mathrm{d}x \simeq \sum_{k = 0}^n W_k(\omega) \cdot f(X_k) \ ,
\end{align*}
where $W_k$ are weights depending on the oscillation frequency $\omega$, given by
\begin{align}
\label{eq:BV_weights}
    W_k(\omega) = \sum_{\ell=0}^n e^{i\omega c}\sqrt{\frac{m \pi}{2\omega}} \,  i^\ell (2\ell + 1) \, J_{\ell+\frac{1}{2}}(m \omega) \, w_k P_\ell(x_k) \ ,
\end{align}
$m = \frac{b-a}{2}$, $c = \frac{b+a}{2}$, and $x_k, w_k$ are the $(n+1)-$points Gauss-Legendre quadrature nodes and weights, and $X_k$ are 
\begin{align*}
    X_k = \frac{b-a}{2} x_k + \frac{b+a}{2} \ .
\end{align*}

\section{Bound on the cutoff $b$}
\label{appendix:cutoff}

In this section we analyze how the choice of cutoff $b$ in the $\zeta$ integration affects the error of the integral in each case. Then, one can use these relations to choose an appropriate $b$ for a desired error tolerance.

We will use the following result: the source function for each block \eqref{eq:source_block} can be bounded by
\begin{align*}
    \lvert s_k(\zeta, n \Delta \tilde{t}) \rvert \leq \sum_{i=n-N_k+1}^{n-N_{k-1}} \lvert q(i \Delta \tilde{t})\rvert  e^{-\zeta^2 \left( n - i \right) \Delta \tilde{t}} \leq Q \ 
    \frac{e^{-\zeta^2 \Delta\tilde{t}(N_{k-1}-1)}-e^{-\zeta^2 \Delta\tilde{t}(N_{k}-1)}}{1 - e^{-\zeta^2 \Delta\tilde{t}}}\ ,
\end{align*}
where $Q = \max_i \ \lvert q(i \Delta \tilde{t})\rvert $.

We will also make use of the following integrals
\begin{align*}
    \int_b^\infty \frac{e^{- \alpha x^2}}{x}  \mathrm{d}x &= \frac{1}{2} \Gamma(0, b^2  \alpha )  \ , \\
    \int_0^b \frac{e^{- \alpha x^2} - e^{- \beta x^2}}{x} \mathrm{d}x &= \frac{1}{2} \left( \Gamma(0, b^2  \beta ) - \Gamma(0, b^2  \alpha ) + \ln \frac{\beta}{\alpha} \right) \ ,
\end{align*}
where $\Gamma(s, x)$ is the incomplete gamma function.
Note that in the last integral, the integral of each term separately does not exist. 

\subsection{Point to point}
The error we commit in cutting off the $\zeta$ integral at $b$ is
\begin{align*}
    \epsilon &= \frac{r_b^2}{\alpha k_g}\Biggr\lvert  \int_0^\infty s_k(\zeta, n \Delta \tilde{t}) \left(1 -  e^{-\zeta^2 \Delta \tilde{t}}\right) \frac{v^\text{P2P}(\zeta)}{\zeta^2} \mathrm{d}\zeta -  \int_0^b s_k(\zeta, n \Delta \tilde{t}) \left(1 -  e^{-\zeta^2 \Delta \tilde{t}}\right) \frac{v^\text{P2P}_k(\zeta)}{\zeta^2} \ \mathrm{d}\zeta \Biggr\rvert \\
    &\leq  \frac{r_b^2}{\alpha k_g}
    Q\int_b^\infty  \left(e^{-\zeta^2 \Delta\tilde{t}(N_{k-1}-1)}-e^{-\zeta^2 \Delta\tilde{t}(N_{k}-1)}\right) \frac{\lvert v^\text{P2P}(\zeta)\rvert}{\zeta^2} \mathrm{d}\zeta \\
    & \leq \frac{Q}{2\pi^2 k_g r}  \int_0^b  \frac{e^{-\zeta^2 \Delta \tilde{t} (N_{k-1} - 1)} -  e^{-\zeta^2 \Delta \tilde{t} (N_{k} - 1)}}{\zeta}  \mathrm{d}\zeta  \\
    &\leq  \frac{Q}{4\pi^2 k_g r}  \left( \Gamma(0, b^2 (N_k-1)\Delta \tilde{t}) - \Gamma(0, b^2 (N_{k-1}-1)\Delta \tilde{t}) + \ln \frac{N_k-1}{N_{k-1}-1}\right) \ .
\end{align*}

\subsection{Line to point}
We will need a bound on the geometrical factor of the line to point case:
\begin{align*}
    \frac{\lvert v^\text{L2P}(\zeta)\rvert }{\zeta^2}&= \frac{\alpha}{2 \pi^2 r_b^2 } \Biggr\lvert  \int_D^{D+H}\frac{\sin{\tilde{r}(z^\prime) \zeta}}{r(z^\prime) \zeta} \mathrm{d}z^\prime \Biggr\rvert \leq   \frac{\alpha}{2 \pi^2 r_b^2 } \int_D^{D+H}\frac{ \lvert\sin{\tilde{r}(z^\prime)\zeta \rvert}}{r(z^\prime) \zeta} \mathrm{d}z^\prime  \\ 
    &\leq  \frac{\alpha}{2 \pi^2 r_b^2 }  \int_D^{D+H}\frac{ 1}{r(z^\prime) \zeta} \mathrm{d}z^\prime = \frac{\alpha}{2 \pi^2 r_b^2 } \frac{1}{\zeta}  \ln{\left(\frac{z-D + \sqrt{\sigma^2 + (z-D)^2}}{z-D-H + \sqrt{\sigma^2 + (z-D-H)^2}}\right)} \ .
\end{align*}

Assume that we can approximate $v^\text{L2P}$ by $ \bar{v}^\text{L2P}_k$ such that
\begin{align*}
    \frac{2 \pi^2 r_b^2 }{\alpha}\frac{\lvert v^\text{L2P}(\zeta) - \bar{v}^\text{L2P}_k(\zeta)\rvert}{\zeta}< \epsilon^\prime \ .
\end{align*}

Then, the error we commit in cutting off the $\zeta$ integral at $b$ is
\begin{align*}
    \epsilon &= \frac{r_b^2}{\alpha k_g}\Biggr\lvert  \int_0^\infty s_k(\zeta, n \Delta \tilde{t}) \left(1 -  e^{-\zeta^2 \Delta \tilde{t}}\right) \frac{v^\text{L2P}(\zeta)}{\zeta^2} \mathrm{d}\zeta -  \int_0^b s_k(\zeta, n \Delta \tilde{t}) \left(1 -  e^{-\zeta^2 \Delta \tilde{t}}\right) \frac{\bar{v}^\text{L2P}_k(\zeta)}{\zeta^2} \ \mathrm{d}\zeta \Biggr\rvert \\
    &\leq 
    Q \frac{r_b^2}{\alpha k_g}\int_0^b  \left(e^{-\zeta^2 \Delta\tilde{t}(N_{k-1}-1)}-e^{-\zeta^2 \Delta\tilde{t}(N_{k}-1)}\right) \frac{\lvert v^\text{L2P}(\zeta) - \bar{v}^\text{L2P}_k(\zeta)\rvert}{\zeta^2} \mathrm{d}\zeta \\
    &\hspace{0.5cm}+ Q\frac{r_b^2}{\alpha k_g}\int_b^\infty   \left(e^{-\zeta^2 \Delta\tilde{t}(N_{k-1}-1)}-e^{-\zeta^2 \Delta\tilde{t}(N_{k}-1)}\right)  \frac{\lvert v^\text{L2P}(\zeta)\rvert }{\zeta^2} \mathrm{d}\zeta \\
    & \leq \frac{Q}{2 \pi^2 k_g} \ \epsilon^\prime \int_0^b  \frac{e^{-\zeta^2 \Delta \tilde{t} (N_{k-1} - 1)} -  e^{-\zeta^2 \Delta \tilde{t} (N_{k} - 1)}}{\zeta}  \mathrm{d}\zeta  \\
    &\hspace{0.5cm}+ \frac{Q}{2 \pi^2 k_g} \ln{\left(\frac{z-D + \sqrt{\sigma^2 + (z-D)^2}}{z-D-H + \sqrt{\sigma^2 + (z-D-H)^2}}\right)} \int_b^\infty  \frac{e^{-\zeta^2 \Delta \tilde{t} (N_{k-1} - 1)} -  e^{-\zeta^2 \Delta \tilde{t} (N_{k} - 1)}}{\zeta}  \mathrm{d}\zeta\\
    &\leq \frac{Q}{4 \pi^2 k_g} \ \epsilon^\prime  \left( \Gamma(0, b^2 (N_k-1)\Delta \tilde{t}) - \Gamma(0, b^2 (N_{k-1}-1)\Delta \tilde{t}) + \ln \frac{N_k-1}{N_{k-1}-1}\right) \\
    & \hspace{0.5cm}+\frac{Q}{4 \pi^2 k_g} \ln{\left(\frac{z-D + \sqrt{\sigma^2 + (z-D)^2}}{z-D-H + \sqrt{\sigma^2 + (z-D-H)^2}}\right)}  \Biggl( \Gamma \left(0, b^2(N_{k-1} - 1)\Delta \tilde{t}\right)-\Gamma \left(0, b^2 (N_{k} - 1)\Delta \tilde{t}\right) \Biggr) \\
    & = \frac{Q}{4 \pi^2 k_g} \Biggl[ \left( \Gamma \left(0, b^2(N_{k-1} - 1)\Delta \tilde{t}\right)- \Gamma \left(0, b^2 (N_{k} - 1)\Delta \tilde{t}\right)\right) \\ &\hspace{0.5cm}\times \left( \ln{\left(\frac{z-D + \sqrt{\sigma^2 + (z-D)^2}}{z-D-H + \sqrt{\sigma^2 + (z-D-H)^2}}\right)} - \epsilon^\prime \right)
    + \epsilon^\prime \ln \frac{N_k-1}{N_{k-1}-1} \Biggr] \ .
\end{align*}

\subsection{Line to line}
We will need a bound on the geometrical factor of the line to line case:
\begin{align*}
    \frac{\lvert v^\text{L2L}(\zeta)\rvert }{\zeta^2}&= \frac{\alpha}{2 \pi^2 r_b^2 }\Biggr\lvert   \int_{D_t}^{D_t+H_t}\int_{D_s}^{D_s+H_s}\frac{\sin{\tilde{r}(z, z^\prime) \zeta}}{r(z, z^\prime) \zeta} \mathrm{d}z^\prime  \mathrm{d}z \Biggr\rvert \leq   \frac{\alpha}{2 \pi^2 r_b^2 }\int_{D_t}^{D_t+H_t}\int_{D_s}^{D_s+H_s}\frac{\lvert \sin{\tilde{r}(z, z^\prime) \zeta} \rvert}{r(z, z^\prime) \zeta} \mathrm{d}z^\prime  \mathrm{d}z \\ 
    &\leq  \frac{\alpha}{2 \pi^2 r_b^2 }\frac{1}{\zeta} \int_{D_t}^{D_t+H_t}\int_{D_s}^{D_s+H_s}\frac{1}{r(z, z^\prime)} \mathrm{d}z^\prime  \mathrm{d}z =\frac{\alpha}{2 \pi^2 r_b^2 } \frac{1}{\zeta}  \left( \beta(d_4) + \beta(d_3) -\beta(d_1)-\beta(d_2)\right) \ ,
\end{align*}
where 
\begin{align*}
    d_1 &= D_s + H_s - D_t \ , \\
    d_2 &= D_s - H_t - D_t \ , \\
    d_3 &= D_s - D_t \ , \\
    d_4 &= D_s + H_s - D_t - H_t \ , \\
    \beta(d) &= \sqrt{\sigma^2 + d^2} + d \log{\left(\sqrt{\sigma^2 + d^2}- d \right)} \ .
\end{align*}

Assume that we can approximate $v^\text{L2L}$ by $ \bar{v}^\text{L2L}_k$ such that
\begin{align*}
    \frac{2 \pi^2 r_b^2 }{\alpha}\frac{\lvert v^\text{L2L}(\zeta) - \bar{v}^\text{L2L}_k(\zeta)\rvert}{\zeta}< \epsilon^\prime \ .
\end{align*}

Then, the error we commit in cutting off the $\zeta$ integral at $b$ is
\begin{align*}
    \epsilon &=  \frac{r_b^2}{\alpha k_g}\Biggr\lvert  \int_0^\infty s_k(\zeta, n \Delta \tilde{t}) \left(1 -  e^{-\zeta^2 \Delta \tilde{t}}\right) \frac{v^\text{L2L}(\zeta)}{\zeta^2} \mathrm{d}\zeta -  \int_0^b s_k(\zeta, n \Delta \tilde{t}) \left(1 -  e^{-\zeta^2 \Delta \tilde{t}}\right) \frac{\bar{v}^\text{L2L}_k(\zeta)}{\zeta^2} \ \mathrm{d}\zeta \Biggr\rvert \\
    &\leq 
    Q  \frac{r_b^2}{\alpha k_g}\int_0^b   \left(e^{-\zeta^2 \Delta\tilde{t}(N_{k-1}-1)}-e^{-\zeta^2 \Delta\tilde{t}(N_{k}-1)}\right) \frac{\lvert v^\text{L2L}(\zeta) - \bar{v}^\text{L2L}_k(\zeta)\rvert}{\zeta^2} \mathrm{d}\zeta \\
    &\hspace{0.5cm}+ Q \frac{r_b^2}{\alpha k_g}\int_b^\infty  \left(e^{-\zeta^2 \Delta\tilde{t}(N_{k-1}-1)}-e^{-\zeta^2 \Delta\tilde{t}(N_{k}-1)}\right)  \frac{\lvert v^\text{L2L}(\zeta)\rvert }{\zeta^2} \mathrm{d}\zeta \\
    & \leq  \frac{Q}{2 \pi^2 k_g} \ \epsilon^\prime \int_0^b  \frac{e^{-\zeta^2 \Delta \tilde{t} (N_{k-1} - 1)} -  e^{-\zeta^2 \Delta \tilde{t} (N_{k} - 1)}}{\zeta}  \mathrm{d}\zeta  \\
    &\hspace{0.5cm}+ \frac{Q}{2 \pi^2 k_g} \left( \beta(d_4) + \beta(d_3) -\beta(d_2)-\beta(d_1)\right) \int_b^\infty  \frac{e^{-\zeta^2 \Delta \tilde{t} (N_{k-1} - 1)} -  e^{-\zeta^2 \Delta \tilde{t} (N_{k} - 1)}}{\zeta}  \mathrm{d}\zeta\\
    &\leq  \frac{Q}{4 \pi^2 k_g}  \epsilon^\prime \left( \Gamma(0, b^2 (N_k-1)\Delta \tilde{t}) - \Gamma(0, b^2 (N_{k-1}-1)\Delta \tilde{t}) + \ln \frac{N_k-1}{N_{k-1}-1}\right) \\
    & \hspace{0.5cm}+ \frac{Q}{4 \pi^2 k_g}\left( \beta(d_4) + \beta(d_3) -\beta(d_1)-\beta(d_2)\right) \Biggl( \Gamma \left(0, b^2(N_{k-1} - 1)\Delta \tilde{t}\right)-\Gamma \left(0, b^2 (N_{k} - 1)\Delta \tilde{t}\right) \Biggr) \\ 
    &=  \frac{Q}{4 \pi^2 k_g} \Biggl( \left( \Gamma \left(0, b^2(N_{k-1} - 1)\Delta \tilde{t}\right)-\Gamma \left(0, b^2 (N_{k} - 1)\Delta \tilde{t}\right)\right) \Big( \beta(d_4) + \beta(d_3) - \beta(d_2) - \beta(d_1) - \epsilon^\prime \Big) \\
    &\hspace{2cm}+  \epsilon^\prime \ln \frac{N_k-1}{N_{k-1}-1} \Biggr) \ .
\end{align*}

\section{Computation of the line integrals}
\label{appendix:mean}

The goal of this section is to derive alternative, easier to integrate  expressions for the computation of the integrals over the line sources. The original expressions are integrals that need to be evaluated piecewise in the different regions that arise due to the geometry. The goal is to rewrite them under a single integral whose integration variable is the physical distance $r$.

\subsection{Line to point}
\label{appendix:line_to_point}
In this case, the relevant integral is
\begin{align*}
    \int_{a}^{b} f(r(z, z^\prime)) \ \mathrm{d}z^\prime \ ,
\end{align*}
where $z$ is considered fixed. In this case, it is enough to perform the change of variable
\begin{align}
\label{eq:change_variable_r}
    r = \sqrt{\sigma^2 + (z-z^\prime)^2} \ ,
\end{align}
with
\begin{align*}
    \mathrm{d} r = -\frac{(z-z^\prime)}{\sqrt{\sigma^2 + (z-z^\prime)^2}} \mathrm{d} z^\prime = \frac{\sqrt{r^2 - \sigma^2}}{r} \mathrm{d} z^\prime \ .
\end{align*}
If $a < z < b$, the change of variable \eqref{eq:change_variable_r} is not bijective, so the integration interval must be split accordingly.
By doing this, we get
\begin{align*}
    \int_{a}^{b} f(r(z, z^\prime)) \ \mathrm{d}z^\prime = \int_{r_1}^{r_3} f(r) \frac{r \ \alpha_L(r)}{\sqrt{r^2 - \sigma^2}}\mathrm{d}r \ ,
\end{align*}
where
\begin{align}
\label{eq:l2p_r_alpha}
\begin{split}
    \alpha_L(r) = 
    \begin{cases}
         2 \ , \ &\text{if} \ \ r_1 < r < r_2 \\
         1 \ , \ &\text{if} \ \ r_2 < r < r_3 \\ 
         0 \ , \ &\text{elsewhere}
    \end{cases}
    \ ,
    \end{split}
\end{align}
and 
\begin{align}
\label{eq:l2p_r_limits}
\begin{split}  
    r_1 &= 
    \begin{cases}
        \sigma \ , \ &\text{if} \ \ a \leq z \leq b \\
        \text{min}\left( \sqrt{\sigma^2 + (z-b)^2}, \sqrt{\sigma^2 + (z-a)^2} \right) \ , \ &\text{otherwise}
    \end{cases} \ , \\
    r_2 &= \text{min}\left( \sqrt{\sigma^2 + (z-b)^2}, \sqrt{\sigma^2 + (z-a)^2} \right) \ , \\ 
    r_3 &= \text{max}\left( \sqrt{\sigma^2 + (z-b)^2}, \sqrt{\sigma^2 + (z-a)^2} \right) \ .
\end{split}
\end{align}

\subsection{Line to line}
\label{appendix:line_to_line}

In this case, the relevant integral is in the rectangle $\Omega_R = [a, b] \times [a^\prime, b^\prime]$:
\begin{align*}
    I_{L2L}(f) = \int_{a^\prime}^{b^\prime} \int_a^b f(r(z, z^\prime)) \ \mathrm{d} z \ \mathrm{d} z^\prime \ .
\end{align*}
It is possible to rewrite $I_{L2L}$ as a single integral in the interval $D$, as stated by the Reisz-Fréchet representation theorem, which implies that there exists a function $h \in L^2(D)$, called the \textit{representative of }$I_{L2L}$, such that
\begin{align*}
     I_{L2L}(f) = \int_D h(r) f(r) \ \mathrm{d}r \ .
\end{align*}
The Reisz-Fréchet theorem does not provide a way to construct such function $h$, only that it can be done. Fortunately, in our case, it is easy to obtain by using the following result:

\begin{theorem}[Coarea formula]\cite{coarea}
Let $\phi: \Omega \rightarrow \mathbb{R}$ a $\mathcal{C}^1(\Omega)$ function with no critical points. Then for any measurable function $f: \Omega \rightarrow \mathbb{R}$, we have
\begin{align}
\label{eq:coarea}
    \int_\Omega f(\mathbf{x}) \ \mathrm{d}V_R(\mathbf{x}) = \int_\mathbb{R} \left( \int_{\phi^{-1}(t)} \frac{f(\mathbf{x})}{\lvert\nabla \phi(\mathbf{x})\rvert} \ \mathrm{d}V_{\phi^{-1}(t)}(\mathbf{x})  \right) \ \mathrm{d}t \ .
\end{align}
\end{theorem}
The interpretation of this formula is that we first integrate the function $f$ in the level sets of the auxiliary function $\phi$, this is, in the subregions $\{x \in D \, | \, \phi(x) = t \}$ for every value of $t$, and then integrate over all the values of $t$. 

Let us choose $r: \Omega \rightarrow \mathbb{R}$, defined by $r(z, z^\prime) = \sqrt{\sigma^2 + (z - z^\prime)^2}$ as the function defining the level sets. Note that the gradient of $r$, $\nabla r$, whose expression in this case is
\begin{align*}
    \nabla r = \frac{\sqrt{r^2 - \sigma^2}}{r} \ \left(1, -1 \right) \ ,
\end{align*}
can be written as a function of $r$. Then, Equation \eqref{eq:coarea} gives
\begin{align*}
    \int_\Omega f(r(\mathbf{x})) \ \mathrm{d}\mathbf{x} =  \int_\mathbb{R} \frac{f(r)}{\lvert\nabla r(r)\rvert} L_\Omega(r)\ \mathrm{d}r \ ,
\end{align*}
where $L_\Omega(r)$ is the induced volume of the level sets of $r$ in $\Omega$.

All that remains is to compute the function $L_{\Omega_R}$ in the particular case of the rectangle $\Omega_R$. 
Note that the level sets of $r$ in $\Omega_R$ consist of segments parallel to $\{ z=z^\prime \}$ so it is enough to compute their length.

The global minimum of $r$, $r_\text{min} = \sigma$, occurs on the line $\{ z=z^\prime \}$. If $\Omega_R$ does not intersect with $\{ z=z^\prime \}$, the minimum value of $r$ in $\Omega_R$ will be determined by the corner of $\Omega_R$ closest to $\{ z=z^\prime \}$. Likewise, the maximum of $r$ will occur in the corner furthest to $\{ z=z^\prime \}$.

Define the disjoint subregions of the rectangle $\Omega_R$
\begin{align*}
    \Omega_R^+ = \Omega_R \cap \{ z < z^\prime \} \ , \ \ \Omega_R^- = \Omega_R \cap \{ z > z^\prime \} \ .
\end{align*}
Then, 
\begin{align*}
    L_{\Omega_R}(r) = \begin{cases}
        L_{\Omega_R^+}(r) + L_{\Omega_R^-}(r) , \ &\text{if} \ r > \sigma  \\
        \max \left( L_{\Omega_R^+}(\sigma), L_{\Omega_R^-}(\sigma) \right) , \ &\text{if} \ r = \sigma 
    \end{cases} \ .
\end{align*}
Note that by the symmetry of the problem, it is enough to compute one of $L_{\Omega_R^+}$ or $L_{\Omega_R^-}$. Indeed, defining the transposed rectangle $\Omega_R^T$ of $\Omega_R$ by
\begin{align*}
    \Omega_R^T = [a^\prime, b^\prime] \times [a, b] \ ,
\end{align*}
it is clear that $L_{\Omega_R^-}(r) = L_{{\Omega_R^T}^+}(r)$ (or alternatively, $L_{\Omega_R^+}(r) = L_{{\Omega_R^T}^-}(r)$).

Let us compute $L_{\Omega_R^+}(r)$. Its functional behaviour is split in several regions determined by $\Omega_R$. The piece-wise expression is easy to compute
\begin{align*}
    L_{\Omega_R^+}(r) = 
    \begin{cases}
        \sqrt{2} \left( b - a^\prime + \sqrt{r^2 - \sigma^2} \right) \ , \ &\text{if} \ \ r^+_1 < r < r^+_2  \\
        \sqrt{2} \ \min (b-a, b^\prime - a^\prime) \ , \ &\text{if} \ \ r^+_2 < r < r^+_3  \\
        \sqrt{2} \left( b^\prime - a  - \sqrt{r^2 - \sigma^2} \right) \ , \ &\text{if} \ \ r^+_3 < r < r^+_4  \\
        0 \ , &\text{elsewhere}
    \end{cases} \ ,
\end{align*}
where the region limits are
\begin{align}
\label{eq:r_i_plus}
\begin{split}    
    r^+_1 &= 
    \begin{cases}
        \sqrt{\sigma^2 + (a^\prime - b)^2} & \text{if} \  \ a^\prime > b \\
        \sigma &\text{if} \ \  a^\prime < b
    \end{cases} \ , \\
    r^+_2 &= 
    \begin{cases}
        \sqrt{\sigma^2 + (a^\prime - a)^2}  &\text{if} \ \ a^\prime > a \ \text{ and } \ b^\prime - a^\prime > b-a \\ 
        \sqrt{\sigma^2 + (b^\prime - b)^2} &\text{if} \ \ a^\prime > a \ \text{ and } \ b^\prime - a^\prime < b-a \\ 
        \sigma  &\text{if} \ \ a^\prime < a
    \end{cases} \ , \\
    r^+_3 &= 
    \begin{cases}
        \sqrt{\sigma^2 + (b^\prime - b)^2} &\text{if} \ \ a^\prime > a \ \text{ and } \ b^\prime - a^\prime > b-a \\ 
        \sqrt{\sigma^2 + (a^\prime - a)^2} &\text{if} \ \ a^\prime > a \ \text{ and } \ b^\prime - a^\prime < b-a \\ 
        \sigma  &\text{if} \ \ a^\prime < a
    \end{cases} \ , \\
    r^+_4 &=
    \begin{cases}
        \sqrt{\sigma^2 + (a - b^\prime)^2} &\text{if} \  \ b^\prime > a  \\
        \sigma &\text{if} \ \  b^\prime < a
    \end{cases} \ ,
\end{split}
\end{align}
given by the values of $r$ where each corner of $\Omega_R^+$ meets the line $\{ z= z^\prime\}$ in increasing order.

With this particular form of $L_{\Omega_R^+}$, we can write the term of the original integral $ I_{L2 L}$ arising from $\Omega^+$ as:
\begin{align*}
    I_{\Omega_R^+}(f) = \int_{r^+_1}^{r^+_4}  f(r) \left(  \frac{r \  \alpha_{\Omega_R^+}(r)}{\sqrt{r^2 - \sigma^2}} + r \ \beta_{\Omega_R^+}(r) \right) \ \mathrm{d}r \ ,
\end{align*}
where
\begin{align}
\label{eq:alpha}
\begin{split}
    &\alpha_{\Omega_R^+}(r) = 
    \begin{cases}
        b - a^\prime &\text{if} \  \ r^+_1 < r < r^+_2 \\
        \min (b- a, b^\prime- a^\prime) &\text{if} \  \ r^+_2 < r < r^+_3 \\
         b^\prime - a  &\text{if} \  \ r^+_3 < r < r^+_4 \\
    \end{cases}, 
\end{split}
        \\
\label{eq:beta}
\begin{split}   
    &\beta_{\Omega_R^+}(r) =
    \begin{cases}
        1 &\text{if} \  \ r^+_1 < r < r^+_2 \\
        0 &\text{if} \  \ r^+_2 < r < r^+_3 \\
        -1 &\text{if} \ \ r^+_3 < r < r^+_4 \\
    \end{cases} 
\end{split} \ .
\end{align}

As discussed, $L_{\Omega_R^-}$ and $r^-_i$ are obtained directly considering the transposed region of $\Omega^+$.
Therefore, $\alpha_{\Omega_R^-}$, $\beta_{\Omega_R^-}$, and $I_{\Omega_R^-}$ are obtained analogously.
Then, by defining $r_1 = \min(r_1^+, r_1^-)$, $r_4 = \max(r_4^+, r_4^-)$, the support of $L_{\Omega_R}$ is $[r_1, r_4]$.
Finally, the integral of interest can be expressed as
\begin{align}
\label{eq:line_to_line_expression}
    I_{L2 L}(f) = \int_{r_1}^{r_4}  f(r) \left(  \frac{r \  \alpha_{\Omega_R}(r)}{\sqrt{r^2 - \sigma^2}} + r \ \beta_{\Omega_R}(r) \right) \ \mathrm{d}r \ ,
\end{align}
where
\begin{align}
\label{eq:alpha_beta}
\begin{split}    
    \alpha_{\Omega_R}(r) = \alpha_{\Omega_R^+}(r)  + \alpha_{\Omega_R^-}(r) \ , \\
    \beta_{\Omega_R}(r) = \beta_{\Omega_R^+}(r)  + \beta_{\Omega_R^-}(r) \ .
\end{split}
\end{align}

Another relevant case is that of the strip defined by
\begin{align*}
   \Omega_S = \Big\{  (z, z^\prime) \ | \ a < z < b, \ \  \ \max \{z - h, \ a^\prime \} < z^\prime < \min \{z + h, \ b^\prime\} \Big\} \ .
\end{align*}
We can obtain the integral over this region from the integral over the rectangle \eqref{eq:line_to_line_expression} by noting that, in terms of $r$, the region can be expressed as
\begin{align*}
   \Omega_S = \Omega_R \cap \{  \sigma \leq r \leq r_h \} \ ,
\end{align*}
where
\begin{align*}
    r_h = \sqrt{\sigma^2 + h^2} \ .
\end{align*}
Therefore, we directly find
\begin{align*}
    L_{\Omega_S}(r) = L_{\Omega_R}(r) \ \mathbf{1}_{\{\sigma \ \leq\ r \ \leq \ r_h\}} \ ,
\end{align*}
which implies that the the expression for the integral remain the same, with the exception of the upper integration limit, which is cut off at $r_h$:
\begin{align}
\label{eq:line_to_line_expression_strip}
    I_{\Omega_S}(f) = \int_{r_1}^{\min\{r4, \ r_h\}}  f(r) \left(  \frac{r \  \alpha_{\Omega_R}(r)}{\sqrt{r^2 - \sigma^2}} + r \ \beta_{\Omega_R}(r) \right) \ \mathrm{d}r \ .
\end{align}

\section{The asymptotic method to compute highly oscillatory integrals}
\label{appendix:asymptotic}
The asymptotic method is a specialized method to compute oscillatory integrals \cite{Iserles_Norsett, olver}, of the type
\begin{align}
\label{eq:asymptotic_integral}
    I = \int_a^b f(x) e^{i \omega g(x)} \ \mathrm{d}x \ .
\end{align}
The method works by integrating by parts \eqref{eq:asymptotic_integral} iteratively. At each iteration $k$, a new term proportional to $\omega^{-k}$ is obtained, generating an infinite series, that is truncated at $n$ terms:
\begin{align}
\label{eq:asymptotic_formula}
    I_n = \sum_{k=1}^n \frac{1}{\left( - i \omega\right)^k} \left( 
    \sigma_k(b) e^{i\omega g(b)} - \sigma_k(a) e^{i\omega g(a)} \right) \ ,
\end{align}
where 
\begin{align}
\label{eq:sigma_functions}
    \sigma_1(x) = \frac{f(x)}{g^\prime(x)} \ , \ \ \sigma_{k+1}(x) = \frac{\sigma_k(x)}{g^\prime(x)} \ .
\end{align}
The error we commit in truncating the series at $n$ terms is:
\begin{align}
\label{eq:asymptotic_error}
    E_n = \frac{1}{(-i\omega)^n} \int_a^b g^\prime(x) \sigma_{n+1}(x) e^{i \omega g(x)} \mathrm{d}x \ .
\end{align}

From these expressions, it is clear that this method gives good results for high frequencies $\omega$.

\end{appendices}

\printbibliography

\end{document}